\documentclass[final,leqno]{siamltex}

\usepackage{amsfonts}
\usepackage{amsmath}
\usepackage{amssymb}
\usepackage{graphicx}
\usepackage{float}
\usepackage{subfigure}
\usepackage{booktabs}
\usepackage{mathrsfs}
\usepackage{multirow}
\usepackage{amssymb, color}
\usepackage{comment}
\usepackage{tikz}
\usetikzlibrary{arrows.meta,positioning,calc}

\providecommand{\email}[1]{\texttt{#1}}
\usepackage{tikz}
\usetikzlibrary{arrows.meta,positioning,calc}

\providecommand{\U}[1]{\protect \rule{.1in}{.1in}}

\newtheorem{assumption}[theorem]{Assumption}
\newtheorem{example}[theorem]{Example}

\title{Robust Nash Equilibria in Competitive Investment and Consumption Games with Relative Performance}

\author{
    Kai Du\thanks{School of Mathematics, Shandong University, Jinan 250100, China. Research supported by  Key R\&D Program of Shandong Province (No.2026CXPT232), Natural Science Foundation of China (No.12571482), Taishan Scholar Project of Shandong Province (No.tsqn202507029).  E-mail: \email{kdu@sdu.edu.cn}.}
    \and Gechun Liang\thanks{Department of Statistics, University of Warwick, Coventry CV4 7AL, U.K. Research
    supported by National Natural Science Foundation of China (No.12571512), Laboratory
of Mathematics for Nonlinear Science, Fudan University and Guangdong Basic and Applied Basic Research Foundation (No.2026A1515011188).
    E-mail:\email{g.liang@warwick.ac.uk}.}
    \and 
    Zhen Wu\thanks{School of Mathematics, State Key Laboratory of Cryptography and Digital Economy Security, Shandong University, 
Jinan 250100, China. Research supported by  National Natural Science Foundation of China (No.12521001, 62561160159), National Key Research and Development Program of China (No.2023YFA1009200), Fundamental and Interdisciplinary Disciplines Breakthrough Plan of the Ministry of Education of China (No. JYB2025XDXM114), Shandong Provincial Key Laboratory of Stochastic System Control and Scientific Computing.
E-mail: \email{wuzhen@sdu.edu.cn}}
}

\begin{document}

\maketitle

\begin{abstract}
We study a robust investment--consumption game in which investors
care about their wealth and consumption relative to the population
averages. Investors trading the same asset face the same uncertain
market coefficients. Under explicit sufficient conditions, we
construct a robust Nash equilibrium with a deterministic and
compatible worst-case parameter profile. We show that competition
can affect the worst-case common volatility: in a multi-asset
market, the lowest admissible volatility can be worst-case because
common volatility affects both an investor's wealth and the
population average. We also construct a robust mean-field
equilibrium and establish the convergence
of the finite-player robust Nash equilibria and the asymptotic 
robust $\epsilon$-Nash equilibrium property of the mean-field strategies. 
\end{abstract}


\begin{AMS}  {91G10, 91A16, 93E20.}  
\end{AMS}

\pagestyle{myheadings} \thispagestyle{plain} \markboth{K. Du, G. Liang, and Z. Wu}{Competitive Investment and Consumption Games}

\section{Introduction}

Institutional investors are frequently evaluated relative to the performance of their peers, benchmarks, or competitors. Such relative performance concerns affect both portfolio allocation and consumption decisions, since an investor may prefer a strategy that performs well compared with the market even when it is not optimal in absolute terms. At the same time, investment decisions are made under substantial uncertainty about expected returns, volatilities, and the dependence structure of asset prices. These market parameters are difficult to estimate accurately and may change over time. It is therefore natural to study investment and consumption decisions in a framework that incorporates both strategic competition and model uncertainty.

The classical investment--consumption problem originates from the work of Merton and has subsequently been extended in many directions. For instance, robust formulations replace the assumption of a known market model by an uncertainty set and evaluate strategies under the  worst-case admissible model. See, for example, \cite{S2008,BP2017,YLZ2019}, which study robust investment or investment--consumption problems under ambiguity in market coefficients. These works, however, focus primarily on a single investor and therefore do not address strategic interactions generated by relative performance concerns.

A separate literature studies portfolio competition under relative performance. In \cite{ET2015}, optimal investment under relative performance concerns is analysed in a continuous-time setting. The finite-player and mean-field formulations were developed further in \cite{LZ2019}, where explicit Nash equilibria are obtained for investors with exponential utility. The inclusion of consumption in  portfolio games was subsequently studied in \cite{LS2020} and \cite{Fu}. Related extensions include competitive and homophilous interactions \cite{HZ}, large heterogeneous populations \cite{TZ}, unbounded controlled common noise \cite{SZ}, competitive investment and hedging under random-measure common noise \cite{BH2024}, jump-diffusion models \cite{LHYZ2025}, non-Markovian markets \cite{FSZ} and \cite{WXZ}, forward utilities \cite{MBC}, and recursive preferences \cite{DRS}. Equilibria with relative performance concerns have also been studied through no-arbitrage arguments and under price impact in \cite{BG2023,BG2024}. These models demonstrate that competition can substantially alter equilibrium risk taking, but they typically assume that the underlying market coefficients are known.

More broadly, robust game theory studies strategic interactions when players face uncertainty about model parameters or payoffs. Foundational contributions include \cite{AB2006}, while distributionally robust Nash and Stackelberg equilibria are considered in \cite{LXYZ2018}. Model uncertainty in general equilibrium is studied in \cite{BR2019}. Robust mean-field games have mainly been developed in linear-quadratic settings. In particular, \cite{HH2017} studies robust mean-field linear-quadratic-Gaussian games with an unknown disturbance, and \cite{MB2017} considers risk-sensitive and robust mean-field games. These formulations provide important general insights, but their linear-quadratic structure does not directly accommodate exponential utility, relative wealth and consumption, or ambiguity in both drift and volatility.

The present paper connects these strands of the literature by studying a robust investment--consumption game under relative performance criteria. Investors have exponential utilities depending on both terminal wealth and intermediate consumption relative to their population averages. They trade risky assets whose drifts, asset-specific volatilities, and common volatilities are uncertain. The ambiguity is represented by compact parameter intervals, and the market selects the  worst-case admissible coefficients. Thus each investor faces two interacting sources of strategic dependence: competition with other investors and adverse selection of the market model.

A distinctive feature of the model is that investors are divided into trading groups. Each investor trades one risky asset, while several investors may trade the same asset. Hence the framework includes both the case in which all investors trade a common asset and the case in which each investor trades a distinct asset. Investors trading the same asset are exposed to the same uncertain coefficients. Therefore, a valid robust Nash equilibrium  requires the individual worst-case parameter selection of all investors in the same group to be compatible. This produces an asset-level compatibility condition that does not arise in standard single agent robust control or in portfolio games with known coefficients.  It also distinguishes our formulation from
the robust investment games in \cite{GJL2026,ZZ2025}, where investors
trading the same assets evaluate their strategies under different
worst-case parameters.

We first analyse the finite-player game. For a fixed admissible strategy profile, we characterize each investor's individual worst-case drift and volatility parameters through the corresponding robust Hamiltonian. The worst-case drift and asset-specific volatility are selected at the endpoints of their ambiguity intervals. The common volatility is more subtle because it affects not only the investor's own risk exposure but also their covariance with the aggregate common noise exposure of the other trading groups. Consequently, the  worst-case common volatility may depend on the strength of competition and need not always be the largest admissible volatility.

To construct an explicit robust Nash equilibrium, we next solve an auxiliary $N$-player investment--consumption game under deterministic market coefficients. For exponential utility, we obtain explicit equilibrium investment strategies and affine wealth-dependent consumption strategies. We then derive sufficient conditions under which the individual worst-case parameter selection is deterministic and coincide across all investors trading the same asset. Under these conditions, the auxiliary Nash equilibrium becomes a robust Nash equilibrium of the original game.

The endogenous selection of the worst-case common volatility leads to a new economic implication. In a single asset economy, relative performance concerns affect equilibrium investment strategies but not the worst-case model: investors continue to regard the largest admissible volatility as least favourable, as in the classical robust Merton problem. In a multi-asset economy, however, sufficiently strong competition may reverse this property. Because investors care about outperforming the population average, a low-common-volatility environment may become more adverse than a high-volatility one. Thus competition can alter not only equilibrium risk taking but also investors' endogenous perception of model uncertainty.

We then study the large-population limit. Mean-field games, introduced independently in \cite{LL2006a,HMC2006} and developed systematically in \cite{CD2018}, provide a tractable approximation for large stochastic games. The connection between finite-player and mean-field systems, including asymptotic robust $\epsilon$-Nash equilibria, is well established in general frameworks; see, for example, \cite{HCM2007,CF2018}. Mean-field methods have also been applied to a wide variety of financial and economic models, including constrained systems \cite{HHN2018}, backward stochastic systems \cite{DHW2018,DHW2019}, and applications such as demand-side management \cite{ACDZ}.

In our setting, the representative investor interacts with the population through groupwise average investment, aggregate wealth, and aggregate consumption. We formulate a robust mean-field control problem in which these quantities are initially treated as exogenous. We derive the optimal investment and consumption strategies together with the worst-case parameter selection, and then impose mean-field consistency conditions. Under explicit assumptions, this system admits a unique solution and yields a closed-form robust mean-field equilibrium.

Finally, we establish the asymptotic relation between the finite-player and mean-field formulations. We show that, as the number of investors tends to infinity and the empirical type distributions converge, the finite-player robust Nash equilibrium  strategies converge to their mean-field counterparts. We further prove that the decentralized mean-field strategies form an asymptotic robust $\epsilon$-Nash equilibrium for the finite-player robust game. This provides a rigorous justification for using the robust mean-field equilibrium as an approximation of the large but finite investment game.

The remainder of the paper is organized as follows. Section~\ref{sec:problem-formulation} introduces the financial market, the uncertainty sets, and the robust Nash equilibrium  concept. Section~\ref{sec:robust-equilibrium} constructs the finite-player robust Nash equilibrium and discusses its economic implications. Section~4 develops the robust mean-field game, derives the consistency conditions, and studies the large-population limit. Section \ref{sec: conclusion} then concludes.


\section{Problem formulation}\label{sec:problem-formulation}

This section introduces the financial market, the investors' optimization
problems, and the notion of robust Nash equilibrium.
The model consists of two coupled components.
On the one hand, investors compete strategically through a relative-performance
criterion, leading to a non-cooperative $N$-player game.
On the other hand, the market coefficients are subject to model uncertainty and
are selected adversarially at the asset level.
The robust Nash equilibrium  combines these two interacting mechanisms.

\subsection{Financial market and parameter uncertainty}
We consider a finite-horizon market over $[0,T]$ defined on a complete filtered probability space
\(
(\Omega,\mathcal F,\mathbb F=(\mathcal F_t)_{0\le t\le T},\mathbb P),
\)
supporting one common Brownian motion
$B$
and independent asset-specific Brownian motion
$(W^1,\ldots,W^K)$.
The financial market consists of one risk-free asset with zero interest rate and
$K$ risky assets.
For each $k=1,\ldots,K$, the price process satisfies
\[
\frac{dS_t^k}{S_t^k}
=
\mu_t^kdt
+
\nu_t^kdW_t^k
+
\sigma_t^kdB_t.
\]
Here
$\mu^k$ denotes the drift,
$\nu^k$ the asset-specific volatility,
and
$\sigma^k$ the common volatility.
Throughout the paper the market coefficients are uncertain.
Rather than assuming a reference probability model,
we assume that the coefficients belong to the uncertainty set.
For each risky asset $k=1,\ldots,K$, we assume that the market coefficients are not known exactly. Instead, the drift and volatility processes are only known to belong to an uncertainty set. More precisely, define
\[
\mathcal U_k^p
:=
\left\{
\vartheta^k=(\mu^k,\nu^k,\sigma^k):
\begin{array}{l}
\mu^k,\nu^k,\sigma^k \text{ are }\mathbb F\text{-progressively measurable},\\[1mm]
\mu_t^k\in[\underline\mu_k,\overline\mu_k],
\nu_t^k\in[\underline\nu_k,\overline\nu_k],
\sigma_t^k\in[\underline\sigma_k,\overline\sigma_k], 0\le t\le T
\end{array}
\right\}.
\]
Here
\(
0<\underline\mu_k\le\overline\mu_k,
\
0\le\underline\nu_k\le\overline\nu_k,
\
0\le\underline\sigma_k\le\overline\sigma_k
\)
are constants.
The widths of the intervals
quantify the level of ambiguity in the drift, asset-specific volatility, and common volatility of the risky asset $k$, respectively.
The overall uncertainty set for the financial market is given by the Cartesian product
\(
\mathcal U^p
:=
\prod_{k=1}^K
\mathcal U_k^p.
\)

\subsection{Investors and trading groups}

We consider $N$ investors competing over the $K$ risky assets with $N\geq K$. Each investor is assigned to a single risky asset throughout the investment
horizon.
More precisely,
for every
$i\in\{1,\ldots,N\}$,
let the assignment map 
\[
\iota_i
\in
\{1,\ldots,K\}
\]
denote the asset traded by investor $i$. This assumption models situations in which portfolio managers specialize in a
given asset class or trading desk while allowing several investors to compete
over the same asset.
It also provides an intermediate framework between two classical settings: (1) $K=N$:
every investor trades a different risky asset; (2)
$K=1$:
all investors compete over a common risky asset.

The assignment map naturally partitions the investors into $K$ disjoint
trading groups. For each asset $k$, define
\[
\mathcal I_k
=
\{i\in\{1,\ldots,N\}:\iota_i=k\},
\]
so that
\(
\sum_{k=1}^K |\mathcal I_k|=N\) and
\(p_k^{(N)}=\frac{|\mathcal I_k|}{N}.
\)
The partition
$\{\mathcal I_1,\ldots,\mathcal I_K\}$
plays a fundamental role throughout the paper.

The model has the following interpretation. Within a fixed group
$\mathcal I_k$, all investors trade the same risky asset $k$ and are
exposed to the same uncertain market coefficients
\(
\vartheta^k=(\mu^k,\nu^k,\sigma^k).
\)
Hence, each group is a common-asset investment game in which several
investors compete while facing the same asset-level uncertainty. Across
groups, investors trade different risky assets, so the full economy is a
multi-asset game composed of these common-asset subgames. In particular, the robust
parameter selection is performed at the asset level: the same
parameter profile $\vartheta^k$ must be compatible with the worst-case
responses of all investors in $\mathcal I_k$.


\subsection{Investment and consumption}

Investor $i$ chooses an investment-consumption strategy
$(\pi_t^i,c_t^i)$, where $\pi_t^i$ denotes the amount invested in the
risky asset $S^{\iota_i}$ and $c_t^i$ denotes the consumption.
The corresponding wealth process satisfies
\[
dX_t^i
=
\pi_t^i
\bigl(
\mu_t^{\iota_i}dt
+
\nu_t^{\iota_i}dW_t^{\iota_i}
+
\sigma_t^{\iota_i}dB_t
\bigr)
-
\kappa c_t^i\,dt,
\qquad
X_0^i=x_i,
\]
where $\kappa\in\{0,1\}$ distinguishes the pure investment problem
($\kappa=0$) from the investment--consumption problem ($\kappa=1$). We allow the consumption process $c^i$ to take negative values.
When $c_t^i<0$, the term $-\kappa c_t^i\,dt$ represents an
inflow of funds into investor $i$'s wealth. For related discussions, see \cite{CDZ2012finite}.

Motivated by the explicit Nash equilibrium characterized later (see
Theorem \ref{thm:fixed-market-equilibrium} below and also \cite{LS2020,LZ2019}), we restrict attention to deterministic
investment strategies and affine wealth-dependent consumption
strategies. Such a class is sufficiently rich to contain the robust
Nash equilibrium while preserving the Markovian structure of the 
state processes used in the dynamic programming principle.
More precisely, an admissible strategy profile is a collection
$(\pi,c)=\{(\pi^i,c^i)\}_{i=1}^N$ satisfying
\[
\pi_t^i=p_i(t),
\qquad
c_t^i=q_i(t)+a(t)X_t^i,
\qquad
i=1,\ldots,N,
\]
where $p_i,q_i\in C([0,T];\mathbb R)$ are investor-specific deterministic
functions and $a\in C([0,T];\mathbb{R})$ is a deterministic function common
to all investors. We denote by $\mathcal U$ the collection of all such
strategy profiles $\mathcal U:=\prod_{i=1}^N\mathcal U_i$ where 
\[
\mathcal U_i
:=
\left\{
(\pi^i,c^i):
\begin{array}{l}
\pi_t^i=p_i(t),\quad
c_t^i=q_i(t)+a(t)X_t^i,\\[1mm]
p_i,q_i, a\in C([0,T];\mathbb R)
\end{array}
\right\}.
\]

The requirement that all investors use the same wealth coefficient in
their affine consumption strategies is motivated by the explicit
equilibrium structure obtained in the model without parameter
uncertainty; see \cite{LS2020}. In that setting, the common coefficient
arises endogenously from the exponential utility specification, the common investment
horizon, and the linear wealth dynamics, whereas investor heterogeneity
affects the intercepts of the consumption strategies.

We retain this common-slope structure in the present model because it is
also essential for the Markovian treatment of model uncertainty. Indeed,
when the consumption strategies of the other investors have the affine form
\(
c_t^j=q_j(t)+a(t)X_t^j,\ j\neq i,
\)
their aggregate consumption can be written as an affine function of 
\(
\widehat X_t^{\,i}
=
\frac1N\sum_{j\ne i}X_t^j.
\)
Consequently, when investor $i$ evaluates the  worst-case market
parameters of the asset that they trade, the relevant state processes
are
\(
\bigl(X^i,\widehat X^{\,i}\bigr),
\)
namely, the investor's own wealth and the average wealth of the remaining
investors. This yields a two-dimensional Markovian formulation of the
individual worst-case parameter-selection problem developed in
Section~\ref{subsec:individual-worst-case}.

Each investor $i$ evaluates both terminal wealth and consumption
relative to the population average through the exponential utility function
\[
U_i(x,m;\gamma_i,\theta_i)
:=
-
\exp\!\left(
-\gamma_i(x-\theta_i m)
\right),
\qquad (x,m)\in\mathbb R^2.
\]
Here $x$ denotes the investor's own wealth (or consumption), while
$m$ represents the corresponding population average.
The parameter $\gamma_i>0$ is the coefficient of absolute risk aversion,
whereas $\theta_i\in[0,1]$ measures the strength of the relative performance concern.
When $\theta_i=0$, the investor only values absolute performance,
while larger values of $\theta_i$ indicate stronger competitive behaviour.

For notational convenience, for any parameters
$\eta=(\eta^1,\ldots,\eta^N)$,
we write
\(
\eta^{-i}
:=
(\eta^1,\ldots,\eta^{i-1},\eta^{i+1},\ldots,\eta^N),
\)
and denote the population average by
\(
\bar\eta
:=
\frac1N\sum_{j=1}^N\eta^j.
\)
In particular, we shall frequently use the average wealth and average consumption processes,
\(
\bar X_t
:=
\frac1N\sum_{j=1}^N X_t^j,
\
\bar c_t
:=
\frac1N\sum_{j=1}^N c_t^j.
\)
Given a parameter profile
\(
\boldsymbol\vartheta
:=
(\vartheta^1,\ldots,\vartheta^K)
\in\mathcal U^p,
\)
and an admissible strategy profile
$(\pi,c)\in\mathcal U$,
the payoff of investor $i$ is defined by
\begin{equation}\label{payoff_functional}
\begin{aligned}
\mathcal J_i
(\pi^i,c^i,\pi^{-i},c^{-i};
\boldsymbol\vartheta)
:=
\mathbb E\Bigg[
&e^{-\delta_iT}
U_i(X_T^i,\bar X_T;\gamma_i,\theta_i)
\\
&+
\kappa\lambda_i
\int_0^T
e^{-\delta_it}
U_i(c_t^i,\bar c_t;\gamma_i,\theta_i)
\,dt
\Bigg].
\end{aligned}
\end{equation}
The parameter $\delta_i\geq0$ represents the discount factor, and the parameter $\lambda_i\geq 0$ determines the relative importance of intermediate
consumption compared with terminal wealth.
The indicator
\(
\kappa\in\{0,1\}
\)
distinguishes the two models considered in this paper:
$\kappa=1$ corresponds to the investment--consumption problem,
whereas $\kappa=0$ reduces to the pure investment problem.

The dependence of the utility function on the population averages
$\bar X$ and $\bar c$
couples the optimization problems of all investors.
Indeed, the payoff of investor $i$ depends not only on their own strategy
$(\pi^i,c^i)$,
but also on the strategies of every other investor through the average wealth
and average consumption.
Consequently, the optimization problem forms a non-cooperative $N$-player
game.

\subsection{Compatible worst-case parameter selection}
\label{subsec:assetwise-worst-case}

Fix an admissible strategy profile $(\pi,c)\in\mathcal U$. The
uncertain market coefficients are asset-specific. In particular, all
investors belonging to the same trading group $\mathcal I_k$ are exposed
to the same parameter profile
\(
\vartheta^k=(\mu^k,\nu^k,\sigma^k)\in\mathcal U_k^p
\).

\begin{definition}[Compatible worst-case parameter profile]
\label{def:common-worst-case}
Fix $(\pi,c)\in\mathcal U$. A parameter profile
\(
\boldsymbol\vartheta^*
=
(\vartheta^{1,*},\ldots,\vartheta^{K,*})
\in\mathcal U^p
\)
is called a compatible worst-case parameter profile
associated with $(\pi,c)$ if,  for every $k=1,\ldots,K$, every $i\in\mathcal I_k$, and
every $\vartheta^k\in\mathcal U_k^p$,
\[
\mathcal J_i
\bigl(
\pi,c;\vartheta^{k,*},\boldsymbol\vartheta^{-k,*}
\bigr)
\le
\mathcal J_i
\bigl(
\pi,c;\vartheta^k,\boldsymbol\vartheta^{-k,*}
\bigr).
\]
\end{definition}

Although the individual investors may have
different worst-case correspondences, all investors trading asset $k$
face the same market parameter $\vartheta^k$. Hence, a valid market
parameter must be simultaneously worst-case for every investor in
$\mathcal I_k$, conditional on the parameters of the other assets.
In general, it does
not imply that
\(
\boldsymbol\vartheta^*
\in
\operatorname*{arg\,min}_{\boldsymbol\vartheta\in\mathcal U^p}
\mathcal J_i(\pi,c;\boldsymbol\vartheta)
\)
for every investor $i$. It only requires assetwise
worst-case optimality with the parameters of the remaining assets held
fixed.

For a fixed admissible strategy profile $(\pi,c)$, define the set-valued
parameter-selection correspondence by
\[
\Gamma(\pi,c)
:=
\left\{
\boldsymbol\vartheta^*\in\mathcal U^p:
\boldsymbol\vartheta^*
\text{ satisfies Definition~\ref{def:common-worst-case}}
\right\}.
\]
If $\Gamma(\pi,c)$ is a singleton for every $(\pi,c)$ under
consideration, we identify it with the corresponding single-valued
mapping and write
\(
\Gamma:\mathcal U\rightarrow\mathcal U^p\),
\(
\Gamma(\pi,c)=\boldsymbol\vartheta^*
\). 
The explicit characterization of $\Gamma$, together with sufficient
conditions for its nonemptiness and uniqueness, is established in
Section~\ref{sec:robust-equilibrium}.

\subsection{Robust Nash equilibrium}
\label{subsec:robust-nash}

A robust Nash equilibrium consists of a Nash equilibrium under a fixed
market model together with a compatible worst-case
parameter profile for the resulting equilibrium strategies. Thus, the
market model is endogenous in the equilibrium construction, but it is
held fixed when a unilateral deviation is evaluated.

\begin{definition}[Robust Nash equilibrium]
\label{def:common-robust-nash}
A pair
\(
\bigl((\pi^*,c^*),\boldsymbol\vartheta^*\bigr)
\in
\mathcal U\times\mathcal U^p
\)
is called a robust Nash equilibrium if:

\begin{enumerate}
\item[(i)] the selected parameter profile is compatible with the
equilibrium strategies,
\[
\boldsymbol\vartheta^*
\in
\Gamma(\pi^*,c^*);
\]

\item[(ii)] conditional on the fixed parameter profile
$\boldsymbol\vartheta^*$, the strategy profile $(\pi^*,c^*)$ is a Nash
equilibrium, namely, for every investor $i=1,\ldots,N$ and every
unilateral deviation $(\pi^i,c^i)\in\mathcal U_i$,
\[
\begin{aligned}
&
\mathcal J_i
\bigl(
\pi^{i,*},c^{i,*},
\pi^{-i,*},c^{-i,*};
\boldsymbol\vartheta^*
\bigr)
\\
&\qquad\ge
\mathcal J_i
\bigl(
\pi^i,c^i,
\pi^{-i,*},c^{-i,*};
\boldsymbol\vartheta^*
\bigr).
\end{aligned}
\]
\end{enumerate}
\end{definition}

Conditions (i) and (ii) can be expressed by
the following chain inequalities: for every \(k\), every \(i\in\mathcal I_k\),
every unilateral deviation \((\pi^i,c^i)\in\mathcal U_i\), and every
\(\vartheta^k\in\mathcal U_k^p\),
\[
\begin{aligned}
\mathcal J_i
\bigl(
\pi^i,c^i,\pi^{-i,*},c^{-i,*};
\boldsymbol\vartheta^*
\bigr)
&\le
\mathcal J_i
\bigl(
\pi^*,c^*;
\boldsymbol\vartheta^*
\bigr)
\\
&\le
\mathcal J_i
\bigl(
\pi^*,c^*;
\vartheta^k,\boldsymbol\vartheta^{-k,*}
\bigr).
\end{aligned}
\]
For each investor \(i\in\mathcal I_k\), this has the familiar
saddle-point form in the pair \(((\pi^i,c^i),\vartheta^k)\), with the
opponents' strategies and the other assets' parameters held fixed. The full
equilibrium is not a single zero-sum saddle point: it combines
the investors' Nash equilibrium conditions with a compatible worst-case parameter
selection that must satisfy the separate minimization conditions of every
investor trading that asset.

The construction is easiest to understand first in a single investor
problem with one risky asset. Although the admissible parameter process
$\vartheta^1\in\mathcal U_1^p$ may be random, we begin by proposing a
deterministic candidate $\vartheta^{1,*}$ and solving the
investment--consumption problem under that fixed market model. This
produces a candidate strategy $(\pi^{1,*},c^{1,*})$ and its
fixed-market value function. The decisive step is then to hold the
strategy fixed and prove, using this value function, that
$\vartheta^{1,*}$ is worst-case among all admissible parameter
processes. If this check succeeds, fixed-market optimality and
worst-case optimality give the saddle-point inequalities
\[
\mathcal J_1(\pi^1,c^1;\vartheta^{1,*})
\leq
\mathcal J_1(\pi^{1,*},c^{1,*};\vartheta^{1,*})
\leq
\mathcal J_1(\pi^{1,*},c^{1,*};\vartheta^1).
\]
The central contribution of the paper is to identify
explicit conditions under which this check succeeds in the
coupled $N$-investor game. In particular,
Assumption~\ref{ass:deterministic-worst-case} gives sufficient
conditions which, after the fixed-market Nash equilibrium  strategies are
substituted into the individual switching conditions, make the
worst-case parameter selection independent of the state variables and
identical for all investors in each $\mathcal I_k$. It thereby
produces a deterministic parameter profile
$\boldsymbol\vartheta^*\in\Gamma(\pi^*,c^*)$, despite the random
parameter processes permitted by $\mathcal U^p$. The three steps
below establish these individual parameter selection, solve the fixed-market
Nash equilibrium  game, and verify their compatibility at the candidate
equilibrium. 

\medskip

\noindent
\textit{Step 1. Individual worst-case parameter selection.}

Fix an admissible strategy profile $(\pi,c)\in\mathcal U$.
For each investor $i\in\mathcal I_k$, hold the parameter profile
$\boldsymbol\vartheta^{-k}$ of the other assets fixed and deterministic,
and characterize the minimizers of
\(
\inf_{\vartheta^k\in\mathcal U_k^p}
\mathcal J_i(\pi,c;\vartheta^k,\boldsymbol\vartheta^{-k}).
\)
The minimizing selection may depend on
$(X^i,\widehat X^{\,i})$ through the derivatives of investor $i$'s
value function. Investors trading the same asset need not select
the same parameters; their compatibility is checked in Step~3.
\medskip

\noindent
\textit{Step 2. Auxiliary Nash equilibrium under deterministic market coefficients.}

For an arbitrary \textit{deterministic} parameter profile
\(
\boldsymbol\vartheta\in\mathcal U^p
\),
consider the corresponding auxiliary $N$-player
investment--consumption game. We derive its Nash equilibrium,
denoted by
\(
(\pi^{\boldsymbol\vartheta},c^{\boldsymbol\vartheta}),
\)
as a function of the prescribed deterministic market coefficients. At
this stage, no claim is made that $\boldsymbol\vartheta$ is worst-case
for the resulting strategy profile.

\medskip

\noindent
\textit{Step 3. Compatibility of the worst-case parameter selection.}

We seek a pair
\(
\bigl((\pi^*,c^*),\boldsymbol\vartheta^*\bigr)
\)
such that
\(
(\pi^*,c^*)
=
(\pi^{\boldsymbol\vartheta^*},
c^{\boldsymbol\vartheta^*})\)
and
\(
\boldsymbol\vartheta^*
\in
\Gamma(\pi^*,c^*).
\)
Thus, the strategies must be a Nash equilibrium for the selected market
parameters, while the selected parameters must be compatible with the
worst-case responses generated by those strategies. To obtain such a pair, we substitute the auxiliary Nash equilibrium strategies
into the individual switching conditions derived in Step~1. Under
suitable sufficient conditions, these switching conditions become deterministic and
independent of the state variables \((X^i,\widehat X^{\,i})\), and select the same deterministic
parameter profile for all investors trading a given asset. The resulting
pair then satisfies both conditions in
Definition~\ref{def:common-robust-nash}.

\medskip

\begin{figure}[h]
\centering
\begin{tikzpicture}[
    >=Latex,
    node distance=8mm,
    box/.style={
        draw=black,
        rounded corners=2pt,
        align=center,
        text width=6.3cm,
        inner sep=5pt,
        fill=gray!7,
        font=\footnotesize
    },
    fixedpoint/.style={
        draw=black,
        double,
        rounded corners=2pt,
        align=center,
        text width=10.5cm,
        inner sep=6pt,
        fill=white,
        font=\footnotesize
    },
    arrow/.style={
        ->,
        thick,
        shorten >=2pt,
        shorten <=2pt
    }
]

\node[box] (responses) {
    \textbf{Step 1. Individual worst-case responses}\\[1mm]
    For fixed $(\pi,c)$ and deterministic
    $\boldsymbol\vartheta^{-k}$, characterize
    the worst-case $\vartheta^k$ for each $i\in\mathcal I_k$.
};

\node[box, right=8mm of responses] (fixedmarket) {
    \textbf{Step 2. Fixed-market Nash equilibrium}\\[1mm]
    For fixed deterministic $\boldsymbol\vartheta$, compute
    $(\pi^{\boldsymbol\vartheta},
      c^{\boldsymbol\vartheta})$.
};

\coordinate (mid) at
    ($(responses.south)!0.5!(fixedmarket.south)$);

\node[fixedpoint, below=12mm of mid] (consistency) {
    \textbf{Step 3. Compatibility and robust Nash equilibrium}\\[1mm]
    Find $((\pi^*,c^*),\boldsymbol\vartheta^*)$ such that
    \(
      (\pi^*,c^*)=
      (\pi^{\boldsymbol\vartheta^*},
       c^{\boldsymbol\vartheta^*})\quad \text{and}
\quad\boldsymbol\vartheta^*\in\Gamma(\pi^*,c^*).
    \)
};

\draw[arrow] (responses.south) -- (consistency.north west);
\draw[arrow] (fixedmarket.south) -- (consistency.north east);

\end{tikzpicture}
\caption{The fixed point couples the fixed-market Nash equilibrium strategies with the
compatible worst-case parameter selection. }
\label{fig:robust-equilibrium-feedback}
\end{figure}
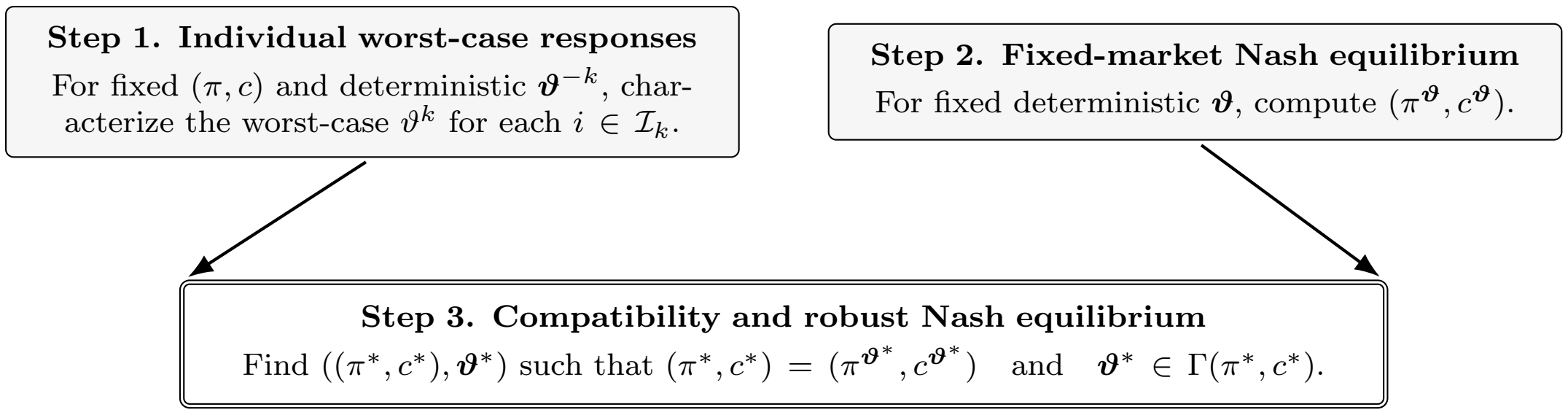

Figure~\ref{fig:robust-equilibrium-feedback} summarizes how the individual worst-case responses and the fixed-market Nash equilibrium  strategies are linked by the compatibility condition.


\section{Construction of the robust Nash equilibrium}
\label{sec:robust-equilibrium}

This section constructs the robust Nash equilibrium  introduced in
Section~\ref{sec:problem-formulation}.
The analysis proceeds in three stages as described above. 


\subsection{Individual worst-case parameter selection}
\label{subsec:individual-worst-case}

Fix an admissible strategy profile $(\pi,c)\in\mathcal U$. For each investor
$i$, we first determine the  worst-case parameters of the asset traded
by that investor while keeping the coefficients of all remaining assets fixed.
This defines an {individual} worst-case model for investor $i$. Since all
investors in the same trading group $\mathcal I_k$ share the same asset
coefficients, these individual minimizers need not coincide. The compatibility
conditions ensuring the existence of a compatible worst-case parameter profile will
be introduced later.

Recall that, for any parameters
$\eta=(\eta^1,\ldots,\eta^N)$, the population average is
\(
\bar\eta
:=
\frac1N
\sum_{j=1}^N
\eta^j.
\)
For a fixed investor $i$, define
\(
\widehat\eta^{\,i}
:=
\frac1N
\sum_{j\neq i}
\eta^j,
\)
so that
\(
\bar\eta
=
\frac1N\eta^i
+
\widehat\eta^{\,i}.
\)
Since
\(
\bar X_t
=
\frac1N X_t^i
+
\widehat X_t^{\,i},
\)
the relative wealth can be written as
\[
X_t^i-\theta_i\bar X_t
=
\left(1-\frac{\theta_i}{N}\right)X_t^i
-
\theta_i\widehat X_t^{\,i}.
\]

For each asset $\ell=1,\ldots,K$, define
\[
\bar\pi_t^\ell
:=
\frac1N
\sum_{j\in\mathcal I_\ell}
\pi_t^j,
\qquad
\widehat\pi_t^{\,i,\iota_i}
:=
\frac1N
\sum_{\substack{j\in\mathcal I_{\iota_i}\\j\neq i}}
\pi_t^j,
\qquad
\widehat c_t^{\,i}
:=
\frac1N
\sum_{j\neq i}
c_t^j.
\]
Then, under the fixed admissible strategy profile $(\pi,c)$, the state
processes $\bigl(X^i,\widehat X^{\,i}\bigr)$ satisfies
\begin{align}
dX_t^i
&=
\pi_t^i
\Bigl(
\mu_t^{\iota_i}\,dt
+
\nu_t^{\iota_i}\,dW_t^{\iota_i}
+
\sigma_t^{\iota_i}\,dB_t
\Bigr)
-
\kappa c_t^i\,dt,
\label{eq:state-individual}
\\
d\widehat X_t^{\,i}
&=
\left[
\sum_{\ell\neq\iota_i}
\mu_t^\ell\bar\pi_t^\ell
+
\mu_t^{\iota_i}\widehat\pi_t^{\,i,\iota_i}
-
\kappa\widehat c_t^{\,i}
\right]dt
\nonumber
+
\left[
\sum_{\ell\neq\iota_i}
\sigma_t^\ell\bar\pi_t^\ell
+
\sigma_t^{\iota_i}\widehat\pi_t^{\,i,\iota_i}
\right]dB_t
\nonumber\\
&\quad
+
\sum_{\ell\neq\iota_i}
\nu_t^\ell\bar\pi_t^\ell\,dW_t^\ell
+
\nu_t^{\iota_i}\widehat\pi_t^{\,i,\iota_i}\,dW_t^{\iota_i}.
\label{eq:state-others}
\end{align}
Since $(\pi,c)\in\mathcal U$, there exist deterministic functions
$p_j,q_j\in C([0,T])$, $j=1,\ldots,N$, and a common function
$a\in C([0,T];\mathbb{R})$ such that
\[
\pi_t^j=p_j(t),
\qquad
c_t^j=q_j(t)+a(t)X_t^j,
\qquad
j=1,\ldots,N.
\]
Consequently, the state equations become
\begin{align*}
dX_t^i
&=
\Bigl(
p_i(t)\mu_t^{\iota_i}
-\kappa q_i(t)
-\kappa a(t)X_t^i
\Bigr)dt
+p_i(t)\nu_t^{\iota_i}\,dW_t^{\iota_i}
+p_i(t)\sigma_t^{\iota_i}\,dB_t,
\\
d\widehat X_t^{\,i}
&=
\Bigg[
\sum_{\ell\ne\iota_i}\mu_t^\ell\bar p^\ell(t)
+\mu_t^{\iota_i}\widehat p^{\,i,\iota_i}(t)
-\kappa\widehat q^{\,i}(t)
-\kappa a(t)\widehat X_t^{\,i}
\Bigg]dt
\\
&\quad+
\Bigg[
\sum_{\ell\ne\iota_i}\sigma_t^\ell\bar p^\ell(t)
+\sigma_t^{\iota_i}\widehat p^{\,i,\iota_i}(t)
\Bigg]dB_t
\\
&\quad+
\sum_{\ell\ne\iota_i}
\nu_t^\ell\bar p^\ell(t)\,dW_t^\ell
+
\nu_t^{\iota_i}\widehat p^{\,i,\iota_i}(t)\,dW_t^{\iota_i}.
\end{align*}
Thus, for a fixed admissible strategy profile, the dynamics of
\((X^i,\widehat X^{\,i})\) form a two-dimensional SDE. This
state reduction does not by itself imply that the pair is Markov:
if the frozen parameter processes of the other assets are arbitrary
\(\mathbb F\)-progressively measurable processes, the coefficients of
this SDE are random and may depend on the past Brownian path.

For the Hamilton-Jacobi-Bellman (HJB) equation below, we impose an additional restriction in
the individual response problem: hold the parameter
profiles $\boldsymbol\vartheta^{-\iota_i}$ of the other assets fixed and
\textit{deterministic}. Under the admissible strategy class, the coefficients in the
affine strategies are deterministic functions of time. Substituting these
strategies and the fixed parameter profiles $\boldsymbol\vartheta^{-\iota_i}$ into
\eqref{eq:state-individual}--\eqref{eq:state-others} shows that the drift
and diffusion coefficients of $(X^i,\widehat X^{\,i})$ depend on time, the
current two-dimensional state, and the control
$\vartheta^{\iota_i}$ only. Thus the system is a two-dimensional
controlled Markov diffusion, and these two state variables suffice for the
dynamic programming principle.

Investor $i$ therefore determines the worst-case parameters of asset
$\iota_i$ by solving
\begin{equation}
\label{eq:individual-robust-problem}
\begin{aligned}
v^i(0,x_i,\widehat x_i;
     \pi,c,\boldsymbol\vartheta^{-\iota_i})
:=
\inf_{\vartheta^{\iota_i}\in\mathcal U_{\iota_i}^p}
\mathbb E\Bigg[
&
e^{-\delta_iT}
U_i\!\left(
\left(1-\frac{\theta_i}{N}\right)X_T^i,
\widehat X_T^{\,i};
\gamma_i,\theta_i
\right)
\\
&
+\kappa\lambda_i
\int_0^T
e^{-\delta_it}
U_i\!\left(
\left(1-\frac{\theta_i}{N}\right)c_t^i,
\widehat c_t^{\,i};
\gamma_i,\theta_i
\right)\,dt
\Bigg],
\end{aligned}
\end{equation}
where
\(
\widehat x_i
:=
\frac1N
\sum_{j\neq i}x_j.
\)

The associated HJB equation is obtained by applying the
dynamic programming principle to the controlled Markov diffusion
\eqref{eq:state-individual}--\eqref{eq:state-others}.
Assuming that the value function
$v^i(t,x,y;\pi,c, \boldsymbol\vartheta^{-\iota_i})$ belongs to
$C^{1,2,2}([0,T)\times\mathbb R^2)\cap C([0,T]\times\mathbb R^2)$, it satisfies the following
HJB equation (\ref{eq:individual-robust-HJB}). For convenience, define switching functions
\begin{align*}
F_\mu^i
&:=
-
\left(
v_x^i\pi_t^i
+
v_y^i\widehat\pi_t^{\,i,\iota_i}
\right),
\\
F_\nu^i
&:=
-
\left[
v_{xx}^i(\pi_t^i)^2
+
2v_{xy}^i\pi_t^i\widehat\pi_t^{\,i,\iota_i}
+
v_{yy}^i(\widehat\pi_t^{\,i,\iota_i})^2
\right],
\\
F_\sigma^i
&:=
\left(
\sum_{\ell\neq\iota_i}
\sigma_t^\ell\bar\pi_t^\ell
\right)
\left(
v_{xy}^i\pi_t^i
+
v_{yy}^i\widehat\pi_t^{\,i,\iota_i}
\right).
\end{align*}
Then the value function $v^i(t,x,y;\pi,c,\boldsymbol\vartheta^{-\iota_i})$ satisfies the HJB equation
\begin{equation}
\label{eq:individual-robust-HJB}
\begin{aligned}
0={}&
v_t^i
-
\sup_{\mu\in[\underline\mu_{\iota_i},\overline\mu_{\iota_i}]}
F_\mu^i\mu
-
\sup_{\nu\in[\underline\nu_{\iota_i},\overline\nu_{\iota_i}]}
\frac12F_\nu^i\nu^2
-
\sup_{\sigma\in[\underline\sigma_{\iota_i},\overline\sigma_{\iota_i}]}
\left(
\frac12F_\nu^i\sigma^2
-
F_\sigma^i\sigma
\right)\\
&
+
v_y^i
\sum_{\ell\neq\iota_i}
\mu_t^\ell\bar\pi_t^\ell
+
\frac12
v_{yy}^i\left(
\sum_{\ell\neq\iota_i}
\sigma_t^\ell\bar\pi_t^\ell
\right)^2
+
\frac12
v_{yy}^i
\sum_{\ell\neq\iota_i}
\left(
\nu_t^\ell\bar\pi_t^\ell
\right)^2
\\
&
-
\kappa
\left[
c_t^iv_x^i
+
\widehat c_t^{\,i}v_y^i
+
\lambda_i
\exp\!\left(
-\delta_it
-
\gamma_i
\left(
1-\frac{\theta_i}{N}
\right)c_t^i
+
\gamma_i\theta_i\widehat c_t^{\,i}
\right)
\right],
\end{aligned}
\end{equation}
subject to the terminal condition
\begin{equation}
\label{eq:individual-robust-terminal}
v^i(T,x,y;\pi,c,\boldsymbol\vartheta^{-\iota_i})
=
-
\exp\!\left(
-\delta_iT
-
\gamma_i
\left(
1-\frac{\theta_i}{N}
\right)x
+
\gamma_i\theta_i y
\right).
\end{equation}
The following proposition characterizes the individual worst-case parameter
selection.

\begin{proposition}
\label{prop:individual-worst-case-response}
Fix $(\pi,c)\in\mathcal U$, an investor
$i\in\{1,\ldots,N\}$, and a deterministic admissible
parameter profile $\boldsymbol\vartheta^{-\iota_i}$
for the other assets.

Then the individual worst-case parameter profile
\(
\vartheta_t^{\iota_i,*}
=
(\mu_t^{\iota_i,*},
\nu_t^{\iota_i,*},
\sigma_t^{\iota_i,*})
\)
is characterized as follows.
The worst-case drift and asset-specific volatility satisfy
\begin{equation}
\label{eq:worst-mu-nu-clean}
\mu_t^{\iota_i,*}
\in
\begin{cases}
\{\overline\mu_{\iota_i}\}, & F_\mu^i>0,\\
[\underline\mu_{\iota_i},\overline\mu_{\iota_i}], & F_\mu^i=0,\\
\{\underline\mu_{\iota_i}\}, & F_\mu^i<0,
\end{cases}
\qquad
\nu_t^{\iota_i,*}
\in
\begin{cases}
\{\overline\nu_{\iota_i}\}, & F_\nu^i>0,\\
[\underline\nu_{\iota_i},\overline\nu_{\iota_i}], & F_\nu^i=0,\\
\{\underline\nu_{\iota_i}\}, & F_\nu^i<0.
\end{cases}
\end{equation}
The worst-case common volatility is a maximizer of
\[
q_i(\sigma)
:=
\frac12F_\nu^i\sigma^2-F_\sigma^i\sigma,
\qquad
\sigma\in
[\underline\sigma_{\iota_i},\overline\sigma_{\iota_i}].
\]
More precisely, the following cases hold.

\begin{enumerate}

\item
If $F_\nu^i<0$, then $q_i$ is strictly concave and
\begin{equation}
\label{eq:worst-sigma-concave}
\sigma_t^{\iota_i,*}
=
\Pi_{[\underline\sigma_{\iota_i},\overline\sigma_{\iota_i}]}
(D_i^\sigma),
\qquad
D_i^\sigma
:=
\frac{F_\sigma^i}{F_\nu^i},
\end{equation}
where
\(
\Pi_{[a,b]}(x)
:=
\min\{b,\max\{a,x\}\}
\)
denotes the Euclidean projection onto $[a,b]$.

\item
If $F_\nu^i=0$, then $q_i(\sigma)=-F_\sigma^i\sigma$ and
\begin{equation}
\label{eq:worst-sigma-linear}
\sigma_t^{\iota_i,*}
\in
\begin{cases}
\{\underline\sigma_{\iota_i}\}, & F_\sigma^i>0,\\
[\underline\sigma_{\iota_i},\overline\sigma_{\iota_i}], & F_\sigma^i=0,\\
\{\overline\sigma_{\iota_i}\}, & F_\sigma^i<0.
\end{cases}
\end{equation}

\item
If $F_\nu^i>0$, then $q_i$ is strictly convex and its maximum over
$[\underline\sigma_{\iota_i},\overline\sigma_{\iota_i}]$
is attained at an endpoint. Specifically,
\begin{equation}
\label{eq:worst-sigma-convex}
\sigma_t^{\iota_i,*}
\in
\begin{cases}
\{\overline\sigma_{\iota_i}\},
&
2D_i^\sigma
<
\underline\sigma_{\iota_i}
+
\overline\sigma_{\iota_i},
\\[1mm]
\{\underline\sigma_{\iota_i}\},
&
2D_i^\sigma
>
\underline\sigma_{\iota_i}
+
\overline\sigma_{\iota_i},
\\[1mm]
\{\underline\sigma_{\iota_i},\overline\sigma_{\iota_i}\},
&
2D_i^\sigma
=
\underline\sigma_{\iota_i}
+
\overline\sigma_{\iota_i}.
\end{cases}
\end{equation}

\end{enumerate}

\end{proposition}

\begin{proof}
The Hamiltonian in \eqref{eq:individual-robust-HJB} is separable with respect
to the three components of
$\vartheta^{\iota_i}=(\mu^{\iota_i},\nu^{\iota_i},\sigma^{\iota_i})$.
Consequently, the optimization over
$\vartheta^{\iota_i}\in\mathcal U_{\iota_i}^p$
reduces to three independent one-dimensional optimization problems.

For the drift, the Hamiltonian is affine in $\mu$, while for the
asset-specific volatility it is affine in $\nu^2$. Therefore,
\eqref{eq:worst-mu-nu-clean} follows immediately.

It remains to maximize
$q_i(\sigma)=\frac12F_\nu^i\sigma^2-F_\sigma^i\sigma$
over the compact interval
$[\underline\sigma_{\iota_i},\overline\sigma_{\iota_i}]$.
When $F_\nu^i\neq0$, completing the square gives
\[
q_i(\sigma)
=
\frac12F_\nu^i(\sigma-D_i^\sigma)^2
-
\frac{(F_\sigma^i)^2}{2F_\nu^i},
\qquad
D_i^\sigma=\frac{F_\sigma^i}{F_\nu^i}.
\]

If $F_\nu^i<0$, then $q_i$ is strictly concave and its constrained maximizer
is the projection of $D_i^\sigma$ onto
$[\underline\sigma_{\iota_i},\overline\sigma_{\iota_i}]$,
proving \eqref{eq:worst-sigma-concave}.

If $F_\nu^i=0$, then $q_i$ is affine, and
\eqref{eq:worst-sigma-linear} follows immediately.

Finally, if $F_\nu^i>0$, then $q_i$ is strictly convex, so its maximum is
attained at one of the endpoints. Moreover,
\[
q_i(\overline\sigma_{\iota_i})
-
q_i(\underline\sigma_{\iota_i})
=
\frac12F_\nu^i
(\overline\sigma_{\iota_i}-\underline\sigma_{\iota_i})
(\underline\sigma_{\iota_i}
+\overline\sigma_{\iota_i}
-2D_i^\sigma).
\]
Comparing the two endpoint values yields
\eqref{eq:worst-sigma-convex}.
\end{proof}
\medskip

The individual worst-case drift and asset-specific volatility are always
attained at the endpoints of their respective uncertainty intervals, except in
the degenerate cases $F_\mu^i=0$ or $F_\nu^i=0$. In contrast, the common
volatility enters the Hamiltonian both through its own quadratic variation and
through its interaction with the aggregate common noise exposure
$\sum_{\ell\neq\iota_i}\sigma_t^\ell\bar\pi_t^\ell$.
Consequently, if $F_\nu^i<0$, the Hamiltonian is concave in $\sigma$, so the
worst-case common volatility may be attained at an interior point of the
admissible interval. If $F_\nu^i>0$, the Hamiltonian is convex, and the
worst-case common volatility is necessarily attained at one of the two
endpoints. This coupling is the main obstacle to obtaining a compatible
worst-case parameter profile in the $N$-player game.

\subsection{Auxiliary Nash equilibrium under deterministic market coefficients}
\label{subsec:fixed-market-equilibrium}

To construct an explicit robust Nash equilibrium, we first consider an auxiliary
investment--consumption game under a fixed deterministic market model.
Throughout this subsection, the parameter profile
\(\boldsymbol\vartheta=(\vartheta^1,\ldots,\vartheta^K)\)
consists of deterministic continuous functions of time. For fixed
opponents' admissible strategies $(\pi^{-i},c^{-i})$, write
\(v^i(t,x,y;\pi^{-i},c^{-i},\boldsymbol\vartheta)\) for investor $i$'s
fixed-market value function. Once the auxiliary Nash equilibrium
has been determined, denote this value function evaluated at the equilibrium
opponents' strategies by \(v^i(t,x,y;\boldsymbol\vartheta)\).

In the next subsection, we identify sufficient conditions under which the
individual worst-case parameter selection is  independent of the state
variables and compatible across all investors trading the same asset. Under
those conditions, the deterministic market coefficients considered here coincide
with the compatible worst-case parameter profile of the original robust game, and the
auxiliary Nash equilibrium constructed below becomes a robust Nash equilibrium.

The following theorem establishes the explicit Nash equilibrium of the
auxiliary game.

\begin{theorem}
\label{thm:fixed-market-equilibrium}
Assume that $\bar\theta<1$ and that, for every $t\in[0,T]$ and
$k=1,\ldots,K$,
\((1-\bar\theta_k)(\nu_t^k)^2+(\sigma_t^k)^2>0.\)
Then the auxiliary $N$-player investment--consumption game associated with
the deterministic market coefficients
$\boldsymbol\vartheta = (\vartheta^1,\ldots,\vartheta^K)$
admits a Nash equilibrium in the admissible class $\mathcal U$.

The equilibrium investment strategy of investor $i$ is
\begin{equation}
\label{eq:equilibrium-investment}
\pi_t^{i,*}
=
\left[
\frac{
\theta_i\sigma_t^{\iota_i}\varphi_t^\pi
}
{
(1-\psi_t^\pi)
\bigl[
(1-\bar\theta_{\iota_i})(\nu_t^{\iota_i})^2
+
(\sigma_t^{\iota_i})^2
\bigr]
}
+
\frac{
\Bigl[
\bigl(
1-\bar\theta_{\iota_i}
+
\gamma_i\theta_i\overline{(1/\gamma)}_{\iota_i}
\bigr)
(\nu_t^{\iota_i})^2
+
(\sigma_t^{\iota_i})^2
\Bigr]
\mu_t^{\iota_i}
}
{
\gamma_i
\bigl[
(1-\bar\theta_{\iota_i})(\nu_t^{\iota_i})^2
+
(\sigma_t^{\iota_i})^2
\bigr]
\bigl[
(\nu_t^{\iota_i})^2
+
(\sigma_t^{\iota_i})^2
\bigr]
}
\right]
\frac1{\Upsilon_t},
\end{equation}
and the equilibrium consumption strategy is
\begin{equation}
\label{eq:equilibrium-consumption}
c_t^{i,*}
=
\kappa
\left[
\frac{\theta_i}{1-\bar\theta}\varphi_t^c
+
\frac{
\ln\lambda_i-\delta_it-\ln\Upsilon_t-f_t^i
}
{\gamma_i}
+
\Upsilon_tX_t^{i,*}
\right],
\end{equation}
where $X^{i,*}$ denotes the corresponding equilibrium wealth process.

The deterministic functions
$\varphi^\pi$, $\psi^\pi$, $\varphi^c$ and $\Upsilon$ are given by
\begin{equation*}
\begin{aligned}
\varphi_t^\pi
&=
\sum_{k=1}^K
\frac{
\overline{(1/\gamma)}_k\mu_t^k\sigma_t^k
}
{
(1-\bar\theta_k)(\nu_t^k)^2+(\sigma_t^k)^2
},
\\
\psi_t^\pi
&=
\sum_{k=1}^K
\frac{
\bar\theta_k(\sigma_t^k)^2
}
{
(1-\bar\theta_k)(\nu_t^k)^2+(\sigma_t^k)^2
},
\\
\varphi_t^c
&=
\frac1N
\sum_{j=1}^N
\frac{
\ln(\lambda_j\Upsilon_t^{-1})
-\delta_jt
-f_t^j
}
{\gamma_j},
\\
\Upsilon_t
&=
\frac{1}{1+\kappa(T-t)},
\end{aligned}
\end{equation*}
where the function $f^i$ is characterized explicitly by the linear ODE (\ref{eq:fixed-market-f-ODE}) below and 
\[
\bar\theta_k
:=
\frac1N\sum_{j\in\mathcal I_k}\theta_j,
\qquad
\overline{(1/\gamma)}_k
:=
\frac1N\sum_{j\in\mathcal I_k}\frac1{\gamma_j},
\qquad
\bar\theta
:=
\frac1N\sum_{j=1}^N\theta_j.
\]
\end{theorem}

\begin{proof}
We divide the proof into three steps.

\noindent
\textit{Step 1.}
Fix a deterministic parameter profile $\boldsymbol\vartheta$ and the
admissible strategies $(\pi^{-i},c^{-i})$ of the other investors. Investor $i$
now chooses their investment and consumption strategies to maximize
their payoff under this prescribed parameter profile. The opponents'
aggregate wealth remains a state variable, and its dynamics are
determined by their fixed admissible strategies. Applying the dynamic programming principle
to $(X^i,\widehat X^{\,i})$ gives the following HJB equation for $v^i(t,x,y;\pi^{-i},c^{-i},\boldsymbol\vartheta)$:
\begin{equation}
\label{eq:HJB-fixed}
\begin{aligned}
0={}&
v_t^i
+
\sup_{\pi}
\Bigg\{
\frac12
\bigl[
(\nu_t^{\iota_i})^2+(\sigma_t^{\iota_i})^2
\bigr]
v_{xx}^i\pi^2
\\
&\qquad
+
\Bigg[
\mu_t^{\iota_i}v_x^i
+
\Bigg(
(\nu_t^{\iota_i})^2\widehat\pi_t^{\,i,\iota_i}
+
\sigma_t^{\iota_i}
\left[
\sum_{\ell\neq\iota_i}
\sigma_t^\ell\bar\pi_t^\ell
+
\sigma_t^{\iota_i}\widehat\pi_t^{\,i,\iota_i}
\right]
\Bigg)v_{xy}^i
\Bigg]\pi
\Bigg\}
\\
&+
\left(
\sum_{\ell\neq\iota_i}
\mu_t^\ell\bar\pi_t^\ell
+
\mu_t^{\iota_i}\widehat\pi_t^{\,i,\iota_i}
-
\kappa\widehat c_t^{\,i}
\right)v_y^i
\\
&+
\frac12
\Bigg[
\left(
\sum_{\ell\neq\iota_i}
\sigma_t^\ell\bar\pi_t^\ell
+
\sigma_t^{\iota_i}\widehat\pi_t^{\,i,\iota_i}
\right)^2
+
\sum_{\ell\neq\iota_i}
(\nu_t^\ell\bar\pi_t^\ell)^2
+
(\nu_t^{\iota_i}\widehat\pi_t^{\,i,\iota_i})^2
\Bigg]v_{yy}^i
\\
&+
\sup_c
\left\{
-\kappa
\left[
cv_x^i
+
\lambda_i
\exp\!\left(
-\delta_it
-\gamma_i\left(1-\frac{\theta_i}{N}\right)c
+\gamma_i\theta_i\widehat c_t^{\,i}
\right)
\right]
\right\},
\end{aligned}
\end{equation}
subject to the terminal condition
\eqref{eq:individual-robust-terminal}. Unlike \eqref{eq:individual-robust-HJB}, which minimizes over the
parameters of asset $\iota_i$ for a fixed admissible strategy profile, this
equation maximizes over investor $i$'s controls while holding all
market parameters fixed.

\noindent
\textit{Step 2.}
Assume temporarily that $v_x^i>0$ and $v_{xx}^i<0$. The Hamiltonian is then
strictly concave with respect to the investment variable, and the
first-order condition gives
\begin{equation}
\label{eq:FOC-pi}
\pi_t^{i,*}
=
-
\frac{
\mu_t^{\iota_i}v_x^i
+
\Bigg(
(\nu_t^{\iota_i})^2\widehat\pi_t^{\,i,\iota_i}
+
\sigma_t^{\iota_i}
\left[
\sum_{\ell\neq\iota_i}
\sigma_t^\ell\bar\pi_t^\ell
+
\sigma_t^{\iota_i}\widehat\pi_t^{\,i,\iota_i}
\right]
\Bigg)v_{xy}^i
}
{
\bigl[
(\nu_t^{\iota_i})^2+(\sigma_t^{\iota_i})^2
\bigr]v_{xx}^i
}.
\end{equation}
\begin{equation}
\label{eq:FOC-c}
c_t^{i,*}
=
\frac{\kappa}{\gamma_i(1-\theta_i/N)}
\left[
\gamma_i\theta_i\widehat c_t^{\,i}
-\delta_it
-
\ln
\frac{v_x^i}
{\lambda_i\gamma_i(1-\theta_i/N)}
\right].
\end{equation}

Motivated by the exponential utility, we seek a solution of the form
\[
v^i(t,x,y;\pi^{-i},c^{-i},\boldsymbol\vartheta)
=
-
\exp
\bigl(
f_t^i+\zeta_t^ix+\xi_t^iy
\bigr).
\]
Substituting this ansatz into \eqref{eq:HJB-fixed} and identifying the
coefficients of $x$, $y$ and the constant term yields
\[
\dot\zeta_t^i
+
\kappa
\frac{(\zeta_t^i)^2}
{\gamma_i(1-\theta_i/N)}
=
0,
\qquad
\zeta_T^i
=
-\gamma_i\left(1-\frac{\theta_i}{N}\right),
\]
and
\[
\dot\xi_t^i
+
\kappa
\frac{\zeta_t^i\xi_t^i}
{\gamma_i(1-\theta_i/N)}
=
0,
\qquad
\xi_T^i
=
\gamma_i\theta_i.
\]
These equations admit the explicit solutions
\[
\zeta_t^i
=
-\gamma_i
\left(
1-\frac{\theta_i}{N}
\right)
\Upsilon_t,
\qquad
\xi_t^i
=
\gamma_i\theta_i\Upsilon_t,
\]
where $\Upsilon_t=\frac{1}{1+\kappa(T-t)}$.

The remaining coefficient satisfies the linear ODE
\begin{equation}
\label{eq:fixed-market-f-ODE}
\dot f_t^i
-
\kappa\Upsilon_tf_t^i
+
g_t^i
=
0,
\qquad
f_T^i
=
-\delta_iT,
\end{equation}
where
\begin{align}\label{g_finite_player}
g_t^i
={}&
\frac{(\gamma_i\theta_i\Upsilon_t)^2}{2}
\Bigg[
\left(
\sum_{\ell\neq\iota_i}
\sigma_t^\ell\bar\pi_t^\ell
\right)^2
+
\sum_{\ell\neq\iota_i}
(\nu_t^\ell\bar\pi_t^\ell)^2
\Bigg]
-
\frac{
\Bigg[
\gamma_i\theta_i
\sigma_t^{\iota_i}
\sum_{\ell\neq\iota_i}
\sigma_t^\ell\bar\pi_t^\ell
\Upsilon_t
+
\mu_t^{\iota_i}
\Bigg]^2
}
{
2\bigl[
(\nu_t^{\iota_i})^2
+
(\sigma_t^{\iota_i})^2
\bigr]
}
\notag\\
&
+
\gamma_i\theta_i\Upsilon_t
\sum_{\ell\neq\iota_i}
\mu_t^\ell\bar\pi_t^\ell
+
\kappa\Upsilon_t
\left(
1+\ln\lambda_i-\delta_it-\ln\Upsilon_t
\right).
\end{align}
The unique solution of \eqref{eq:fixed-market-f-ODE} is
\[
f_t^i
=
\Upsilon_t
\left(
-\delta_iT
+
\int_t^T
\frac{g_s^i}{\Upsilon_s}\,ds
\right).
\]
Consequently,
\(
v^i(t,x,y;\pi^{-i},c^{-i},\boldsymbol\vartheta)
=
-
\exp
\left(
f_t^i
-
\gamma_i
\left(
1-\frac{\theta_i}{N}
\right)
\Upsilon_tx
+
\gamma_i\theta_i\Upsilon_ty
\right).
\)
Differentiating yields $v_x^i>0$ and $v_{xx}^i<0$, thereby verifying the
conditions required for the first-order conditions.

Substituting the explicit derivatives of the value function into
\eqref{eq:FOC-pi}--\eqref{eq:FOC-c} gives
\begin{equation}
\label{eq:investment-best-response}
\pi_t^{i,*}
=
\frac{
\theta_i
\left[
\bigl((\nu_t^{\iota_i})^2+(\sigma_t^{\iota_i})^2\bigr)
\widehat\pi_t^{\,i,\iota_i}
+
\sigma_t^{\iota_i}
\sum_{\ell\ne\iota_i}
\sigma_t^\ell\bar\pi_t^\ell
\right]
+
\mu_t^{\iota_i}/(\gamma_i\Upsilon_t)
}{
(1-\theta_i/N)
\bigl[(\nu_t^{\iota_i})^2+(\sigma_t^{\iota_i})^2\bigr]
}.
\end{equation}
\begin{equation}
\label{eq:consumption-best-response}
c_t^{i,*}
=
\kappa\left[
\Upsilon_tX_t^{i,*}
+
\frac{
\theta_i
\bigl(\widehat c_t^{\,i}
-\Upsilon_t\widehat X_t^{\,i}\bigr)
}{
1-\theta_i/N
}
+
\frac{
\ln\lambda_i-\delta_it-\ln\Upsilon_t-f_t^i
}{
\gamma_i(1-\theta_i/N)
}
\right].
\end{equation}

\medskip
\noindent
\medskip
\noindent
\textit{Step 3.}
At equilibrium, define
\[
\bar\pi_t^{k,*}
:=
\frac1N\sum_{j\in\mathcal I_k}\pi_t^{j,*},
\qquad
\widehat\pi_t^{\,i,\iota_i,*}
:=
\frac1N
\sum_{\substack{j\in\mathcal I_{\iota_i}\\j\ne i}}
\pi_t^{j,*}
=
\bar\pi_t^{\iota_i,*}-\frac1N\pi_t^{i,*}.
\]
Substituting
the above second identity
into \eqref{eq:investment-best-response} and collecting
the terms involving $\pi_t^{i,*}$ gives
\begin{equation}
\label{eq:fixed-market-investment-identity}
\pi_t^{i,*}
=
\frac{
\theta_i
\left[
(\nu_t^{\iota_i})^2\bar\pi_t^{\iota_i,*}
+
\sigma_t^{\iota_i}
\sum_{\ell=1}^K\sigma_t^\ell\bar\pi_t^{\ell,*}
\right]
+
\mu_t^{\iota_i}/(\gamma_i\Upsilon_t)
}{
(\nu_t^{\iota_i})^2+(\sigma_t^{\iota_i})^2
}.
\end{equation}
For each $k$, multiplying
\eqref{eq:fixed-market-investment-identity} by $1/N$,
summing over $i\in\mathcal I_k$, and rearranging yields
\begin{equation}
\label{group_equation}
\bar\pi_t^{k,*}
=
\frac{
\bar\theta_k\sigma_t^k
\displaystyle\sum_{\ell=1}^K
\sigma_t^\ell\bar\pi_t^{\ell,*}
+
\overline{(1/\gamma)}_k\mu_t^k/\Upsilon_t
}{
(1-\bar\theta_k)(\nu_t^k)^2+(\sigma_t^k)^2
}.
\end{equation}
Multiplying \eqref{group_equation} by $\sigma_t^k$ and
summing over $k=1,\ldots,K$ gives
\[
\sum_{k=1}^K\sigma_t^k\bar\pi_t^{k,*}
=
\psi_t^\pi
\sum_{k=1}^K\sigma_t^k\bar\pi_t^{k,*}
+
\frac{\varphi_t^\pi}{\Upsilon_t}.
\]
Since
\(
0\leq\psi_t^\pi
\leq\sum_{k=1}^K\bar\theta_k
=\bar\theta<1,
\)
we obtain 
\begin{equation}
\label{eq:fixed-market-aggregate-investment}
\sum_{k=1}^K\sigma_t^k\bar\pi_t^{k,*}
=
\frac{\varphi_t^\pi}
{(1-\psi_t^\pi)\Upsilon_t}.
\end{equation}
Substituting \eqref{eq:fixed-market-aggregate-investment}
into \eqref{group_equation} determines each
$\bar\pi_t^{k,*}$. Substituting these averages and
\eqref{eq:fixed-market-aggregate-investment} into
\eqref{eq:fixed-market-investment-identity} then gives
\eqref{eq:equilibrium-investment}.
With these investment averages,
\eqref{eq:fixed-market-f-ODE} determines the functions $f^i$.

For consumption, define
\[
\bar X_t^*
:=
\frac1N\sum_{j=1}^N X_t^{j,*},
\qquad
\bar c_t^*
:=
\frac1N\sum_{j=1}^N c_t^{j,*}.
\]
Then
\(
\widehat X_t^{\,i,*}
=
\bar X_t^*-N^{-1}X_t^{i,*}
\)
and
\(
\widehat c_t^{\,i,*}
=
\bar c_t^*-N^{-1}c_t^{i,*}.
\)
For $\kappa=1$, substituting the identities for
$\widehat X_t^{\,i,*}$ and $\widehat c_t^{\,i,*}$
into \eqref{eq:consumption-best-response} and collecting
the terms involving $c_t^{i,*}$ gives
\begin{equation}
\label{eq:fixed-market-consumption-identity}
c_t^{i,*}
=
\theta_i\bar c_t^*
+
\frac{\ln\lambda_i-\delta_it-\ln\Upsilon_t-f_t^i}
{\gamma_i}
+
\Upsilon_t\bigl(X_t^{i,*}-\theta_i\bar X_t^*\bigr).
\end{equation}
Averaging over $i$ therefore yields
\(
(1-\bar\theta)\bar c_t^*
=
\varphi_t^c+(1-\bar\theta)\Upsilon_t\bar X_t^*.
\)
Because $\bar\theta<1$,
\(
\bar c_t^*
=
\Upsilon_t\bar X_t^*
+
\varphi_t^c/(1-\bar\theta).
\)
Substitution into \eqref{eq:fixed-market-consumption-identity}
cancels the terms involving $\bar X_t^*$ and gives
\eqref{eq:equilibrium-consumption} for $\kappa=1$. If $\kappa=0$, consumption is
fixed at zero, and the same stated formula holds because of its
factor $\kappa$.
\end{proof}

\medskip

Theorem~\ref{thm:fixed-market-equilibrium} recovers the equilibrium
results of \cite{LZ2019} as special cases. When $K=1$, all investors
trade a common risky asset. When $K=N$, each investor is assigned to a
different risky asset. The present framework therefore interpolates
between a common-asset investment game and a multi-asset game with
investor-specific assets.

The equilibrium portfolio in
\eqref{eq:equilibrium-investment} has two economically distinct
components. The first component,
\[
\frac{\theta_i\sigma_t^{\iota_i}\varphi_t^\pi}
{(1-\psi_t^\pi)
\left[
(1-\bar\theta_{\iota_i})(\nu_t^{\iota_i})^2
+(\sigma_t^{\iota_i})^2
\right]},
\]
is generated by relative performance concerns. It depends on the
aggregate exposure of all trading groups through $\varphi_t^\pi$ and
$\psi_t^\pi$. Hence, the investment decision of an investor trading
asset $\iota_i$ is affected by the investment opportunities and
characteristics of investors trading other assets.

The second component,
\[
\frac{
\Bigl[
\bigl(
1-\bar\theta_{\iota_i}
+\gamma_i\theta_i\overline{(1/\gamma)}_{\iota_i}
\bigr)(\nu_t^{\iota_i})^2
+(\sigma_t^{\iota_i})^2
\Bigr]\mu_t^{\iota_i}
}{
\gamma_i
\bigl[
(1-\bar\theta_{\iota_i})(\nu_t^{\iota_i})^2
+(\sigma_t^{\iota_i})^2
\bigr]
\bigl[
(\nu_t^{\iota_i})^2+(\sigma_t^{\iota_i})^2
\bigr]
},
\]is the direct risk return motive. It reflects the
expected excess return and the total risk of the asset traded by investor
$i$, adjusted for their own risk tolerance and competition weight. The
characteristics of the investors trading the same asset enter this term
through the population-normalized group quantities
$\bar\theta_{\iota_i}$ and
$\overline{(1/\gamma)}_{\iota_i}$.

Consequently, allowing several investors to trade the same asset has two
effects. It changes the strategic environment within that asset group,
because the group's aggregate competition weight and risk tolerance
affect the common equilibrium exposure. It also changes the aggregate
market exposure entering $\varphi_t^\pi$ and $\psi_t^\pi$, thereby
affecting investors trading other assets. The sign of the resulting
change in an individual portfolio is not generally determined: the
relative performance motive may increase investment in order to keep up
with the benchmark, whereas the associated change in aggregate common
risk may increase the incentive to hedge and reduce investment.


\subsection{Deterministic robust Nash equilibrium}

Proposition~\ref{prop:individual-worst-case-response} describes the
worst-case parameters of the asset traded by investor $i$ for a fixed admissible 
strategy profile $(\pi,c)$ and a deterministic parameter profile
$\boldsymbol\vartheta^{-\iota_i}$ for the other assets. Its switching
functions involve derivatives of
$v^i(t,x,y;\pi,c,\boldsymbol\vartheta^{-\iota_i})$ and may depend on
$(x,y)$. In particular, individual worst-case responses need not agree
among investors trading the same asset.

We now choose a deterministic parameter profile, solve the auxiliary Nash equilibrium 
game under that parameter profile, and verify that its fixed-market value function
also solves the fixed-strategy worst-case problem in Section \ref{subsec:individual-worst-case}. The conditions below
ensure that the proposed parameters minimize the Hamiltonian for every
investor in each trading group. 

\begin{assumption}[Deterministic and compatible worst-case parameter conditions]
\label{ass:deterministic-worst-case}
Assume the conditions of
Theorem~\ref{thm:fixed-market-equilibrium}.
Suppose further that
\(
\underline\sigma_k+\overline\nu_k>0,
\
k=1,\ldots,K,
\)
and that the asset set admits a partition
\[
\{1,\ldots,K\}
=
\mathcal M\cup\overline{\mathcal M},
\qquad
\mathcal M\cap\overline{\mathcal M}
=
\varnothing.
\]

Define
\[
\Phi^{(N)}
:=
\frac{
\displaystyle
\sum_{k\in\mathcal M}
\frac{\overline{(1/\gamma)}_k\underline\mu_k\overline\sigma_k}
{(1-\bar\theta_k)\overline\nu_k^2+\overline\sigma_k^2}
+
\sum_{k\in\overline{\mathcal M}}
\frac{\overline{(1/\gamma)}_k\underline\mu_k\underline\sigma_k}
{(1-\bar\theta_k)\overline\nu_k^2+\underline\sigma_k^2}
}
{
\displaystyle
1
-
\sum_{k\in\mathcal M}
\frac{\bar\theta_k\overline\sigma_k^2}
{(1-\bar\theta_k)\overline\nu_k^2+\overline\sigma_k^2}
-
\sum_{k\in\overline{\mathcal M}}
\frac{\bar\theta_k\underline\sigma_k^2}
{(1-\bar\theta_k)\overline\nu_k^2+\underline\sigma_k^2}
}.
\]

Assume that
\begin{itemize}

\item for every $k\in\mathcal M$,
\[
\max_{i\in\mathcal I_k}\gamma_i\theta_i
\le
\frac{
(\underline\sigma_k+\overline\sigma_k)
\bigl[(1-\bar\theta_k)\overline\nu_k^2+\overline\sigma_k^2\bigr]
\underline\mu_k
}
{
\bigl[
2\overline\nu_k^2
+
\overline\sigma_k(\overline\sigma_k-\underline\sigma_k)
\bigr]
\Bigl(
(1-\bar\theta_k)
(\overline\nu_k^2+\overline\sigma_k^2)
\Phi^{(N)}
-
\overline{(1/\gamma)}_k
\underline\mu_k\overline\sigma_k
\Bigr)
};
\]

\item for every $k\in\overline{\mathcal M}$,
\[
2\overline\nu_k^2
>
(\overline\sigma_k-\underline\sigma_k)\underline\sigma_k,
\]
and
\[
\min_{i\in\mathcal I_k}\gamma_i\theta_i
\ge
\frac{
(\underline\sigma_k+\overline\sigma_k)
\bigl[(1-\bar\theta_k)\overline\nu_k^2+\underline\sigma_k^2\bigr]
\underline\mu_k
}
{
\bigl[
2\overline\nu_k^2
-
\underline\sigma_k(\overline\sigma_k-\underline\sigma_k)
\bigr]
\Bigl(
(1-\bar\theta_k)
(\overline\nu_k^2+\underline\sigma_k^2)
\Phi^{(N)}
-
\overline{(1/\gamma)}_k
\underline\mu_k\underline\sigma_k
\Bigr)
}.
\]

\end{itemize}

\end{assumption}

\begin{theorem}
\label{thm:deterministic-robust-equilibrium}
Suppose Assumption~\ref{ass:deterministic-worst-case}
holds. Define the deterministic parameter profile
$\boldsymbol\vartheta^*$ by
\[
\mu_t^{k,*}=\underline\mu_k,\qquad
\nu_t^{k,*}=\overline\nu_k,\qquad
\sigma_t^{k,*}=\sigma_k^*
=
\begin{cases}
\overline\sigma_k, & k\in\mathcal M,\\[1mm]
\underline\sigma_k, & k\in\overline{\mathcal M},
\end{cases}
\qquad k=1,\ldots,K.
\]
Let $(\pi^*,c^*)$ be the auxiliary Nash equilibrium
\eqref{eq:equilibrium-investment}--
\eqref{eq:equilibrium-consumption} evaluated at
$\boldsymbol\vartheta^*$.

Then $\boldsymbol\vartheta^*\in\Gamma(\pi^*,c^*)$, and
$\bigl((\pi^*,c^*),\boldsymbol\vartheta^*\bigr)$ is a robust
Nash equilibrium. Moreover, \[
v^i(t,x,y;\pi^*,c^*,\boldsymbol\vartheta^{-\iota_i,*})
=
v^i(t,x,y;\boldsymbol\vartheta^*).
\]
Here the left-hand side is the fixed-strategy worst-case value function
defined in Section~\ref{subsec:individual-worst-case}, evaluated at
$(\pi^*,c^*)$ with the parameters of the other assets fixed at
$\boldsymbol\vartheta^{-\iota_i,*}$. The right-hand side is the
fixed-market value function from
Section~\ref{subsec:fixed-market-equilibrium}, with the opponents using
their auxiliary Nash equilibrium strategies under
$\boldsymbol\vartheta^*$.

\end{theorem}

\medskip



\begin{proof} We divide the proof into three steps.

\noindent
\textit{Step 1.}
We begin with the fixed-market value function
\(v^i(t,x,y;\boldsymbol\vartheta^*)\) from
Theorem~\ref{thm:fixed-market-equilibrium}, with the opponents using
\((\pi^{-i,*},c^{-i,*})\). It has the form
\[
v^i(t,x,y;\boldsymbol\vartheta^*)
=
-\exp\!\left(
f_t^i
-\gamma_i\left(1-\frac{\theta_i}{N}\right)\Upsilon_t x
+\gamma_i\theta_i\Upsilon_t y
\right),
\]
where \(f^i\) is determined by
\eqref{eq:fixed-market-f-ODE} evaluated at
\(\boldsymbol\vartheta^*\). In the calculations below,
\(v_x^i,v_y^i\) and the second derivatives denote derivatives of
this value function. Likewise,
\(F_\mu^i,F_\nu^i,F_\sigma^i\) denote the switching functions
defined in Section~\ref{subsec:individual-worst-case} after
substitution of these derivatives and the candidate strategies.

At the candidate deterministic parameter profile, the step 3
in the proof of Theorem~\ref{thm:fixed-market-equilibrium} gives
\[
\sum_{\ell=1}^K
\sigma_\ell^*\bar\pi_t^{\ell,*}
=
\frac{\Phi^{(N)}}{\Upsilon_t},
\qquad
\bar\pi_t^{k,*}
=
\frac{
\bar\theta_k\sigma_k^*\Phi^{(N)}
+\overline{(1/\gamma)}_k\underline\mu_k
}{
\bigl[
(1-\bar\theta_k)\overline\nu_k^2+(\sigma_k^*)^2
\bigr]\Upsilon_t
}.
\]
Because \(\theta_i\geq0\), \(\sigma_k^*\geq0\), and
\(\underline\mu_k>0\), these equations imply
\(\bar\pi_t^{k,*}>0\) for every \(k\). In particular,
\(\sum_{\ell\ne\iota_i}
\sigma_\ell^*\bar\pi_t^{\ell,*}\geq0\).
In turn, 
\[
\pi_t^{i,*}-\theta_i\bar\pi_t^{\iota_i,*}
=
\frac{
\underline\mu_{\iota_i}/\gamma_i
+
\theta_i\sigma_{\iota_i}^*\Upsilon_t
\displaystyle\sum_{\ell\ne\iota_i}
\sigma_\ell^*\bar\pi_t^{\ell,*}
}{
\bigl[
\overline\nu_{\iota_i}^{\,2}
+(\sigma_{\iota_i}^*)^2
\bigr]\Upsilon_t
}
>0.
\]
Since
\(\bar\pi_t^{\iota_i,*}
=N^{-1}\pi_t^{i,*}
+\widehat\pi_t^{\,i,\iota_i,*}\), it follows that
\[
\pi_t^{i,*}
-
\frac{\theta_i}{1-\theta_i/N}
\widehat\pi_t^{\,i,\iota_i,*}
=
\frac{
\pi_t^{i,*}-\theta_i\bar\pi_t^{\iota_i,*}
}{
1-\theta_i/N
}
>0.
\]

Differentiating the value function $v^i(t,x,y;\boldsymbol\vartheta^*)$ yields
\[
v_y^i
=
-\frac{\theta_i}{1-\theta_i/N}v_x^i,
\qquad
v_{xy}^i
=
-\frac{\theta_i}{1-\theta_i/N}v_{xx}^i,
\qquad
v_{yy}^i
=
\left(\frac{\theta_i}{1-\theta_i/N}\right)^2v_{xx}^i,
\]
where \(v_x^i>0\) and \(v_{xx}^i<0\). Substitution into the
switching functions gives, at every \((t,x,y)\),
\[
\begin{aligned}
F_\mu^i
&=
-v_x^i
\left(
\pi_t^{i,*}
-\frac{\theta_i}{1-\theta_i/N}
\widehat\pi_t^{\,i,\iota_i,*}
\right)<0,\\
F_\nu^i
&=
-v_{xx}^i
\left(
\pi_t^{i,*}
-\frac{\theta_i}{1-\theta_i/N}
\widehat\pi_t^{\,i,\iota_i,*}
\right)^2>0,\\
F_\sigma^i
&=
-\frac{\theta_i v_{xx}^i}{1-\theta_i/N}
\left(
\pi_t^{i,*}
-\frac{\theta_i}{1-\theta_i/N}
\widehat\pi_t^{\,i,\iota_i,*}
\right)
\sum_{\ell\ne\iota_i}
\sigma_\ell^*\bar\pi_t^{\ell,*}
\geq0.
\end{aligned}
\]
Thus, for the fixed admissible strategies \((\pi^*,c^*)\), the selected
\(\underline\mu_{\iota_i}\) and
\(\overline\nu_{\iota_i}\) minimize their respective terms in
the Hamiltonian evaluated at the
fixed-market value function $v^i(t,x,y;\boldsymbol\vartheta^*)$.

\noindent
\textit{Step 2.}
It remains to check the selected common volatility. Let
\(k=\iota_i\). Since \(F_\nu^i>0\), the expression
\(\frac12F_\nu^i\sigma^2-F_\sigma^i\sigma\) is convex in
\(\sigma\). Its maximum over
\([\underline\sigma_k,\overline\sigma_k]\) is attained at an
endpoint. Moreover, the derivative identities above give
\[
\frac{F_\sigma^i}{F_\nu^i}
=
\frac{
\theta_i\displaystyle\sum_{\ell\ne k}
\sigma_\ell^*\bar\pi_t^{\ell,*}
}{
\pi_t^{i,*}-\theta_i\bar\pi_t^{k,*}
}.
\]
Consequently, the upper endpoint is a maximizer when
\[
\frac{\underline\sigma_k+\overline\sigma_k}{2}
\bigl(\pi_t^{i,*}-\theta_i\bar\pi_t^{k,*}\bigr)
\geq
\theta_i\sum_{\ell\ne k}
\sigma_\ell^*\bar\pi_t^{\ell,*}.
\]
The lower endpoint is a maximizer when the inequality is reversed.
To check these inequalities at the proposed value $\sigma_k^*$,
evaluate \eqref{group_equation} and
\eqref{eq:fixed-market-aggregate-investment} at
$\boldsymbol\vartheta^*$. Using the definition of $\Phi^{(N)}$
gives
\[
\Upsilon_t\sum_{\ell\ne k}
\sigma_\ell^*\bar\pi_t^{\ell,*}
=
\frac{
(1-\bar\theta_k)
\bigl[\overline\nu_k^2+(\sigma_k^*)^2\bigr]\Phi^{(N)}
-
\overline{(1/\gamma)}_k
\underline\mu_k\sigma_k^*
}{
(1-\bar\theta_k)\overline\nu_k^2+(\sigma_k^*)^2
}.
\]
Similarly, evaluating
\eqref{eq:fixed-market-investment-identity} at
$\boldsymbol\vartheta^*$ and subtracting
$\theta_i\bar\pi_t^{k,*}$ yields
\[
\pi_t^{i,*}-\theta_i\bar\pi_t^{k,*}
=
\frac{
\underline\mu_k/\gamma_i
+
\theta_i\sigma_k^*\Upsilon_t
\displaystyle\sum_{\ell\ne k}
\sigma_\ell^*\bar\pi_t^{\ell,*}
}{
\bigl[\overline\nu_k^2+(\sigma_k^*)^2\bigr]\Upsilon_t
}.
\]

If \(k\in\mathcal M\), substitute
\(\sigma_k^*=\overline\sigma_k\) into these identities.
After multiplying by the positive denominators, the condition
for the upper endpoint becomes
\[
\begin{aligned}
&\gamma_i\theta_i
\bigl[
2\overline\nu_k^2
+\overline\sigma_k
(\overline\sigma_k-\underline\sigma_k)
\bigr]
\Bigl[
(1-\bar\theta_k)
(\overline\nu_k^2+\overline\sigma_k^2)\Phi^{(N)}
-
\overline{(1/\gamma)}_k
\underline\mu_k\overline\sigma_k
\Bigr]
\\
&\qquad\leq
(\underline\sigma_k+\overline\sigma_k)
\bigl[
(1-\bar\theta_k)\overline\nu_k^2
+\overline\sigma_k^2
\bigr]\underline\mu_k.
\end{aligned}
\]
This is the upper-endpoint inequality in
Assumption~\ref{ass:deterministic-worst-case} for investor \(i\).
If \(k\in\overline{\mathcal M}\), substitute
\(\sigma_k^*=\underline\sigma_k\). The condition for the lower
endpoint becomes
\[
\begin{aligned}
&\gamma_i\theta_i
\bigl[
2\overline\nu_k^2
-\underline\sigma_k
(\overline\sigma_k-\underline\sigma_k)
\bigr]
\Bigl[
(1-\bar\theta_k)
(\overline\nu_k^2+\underline\sigma_k^2)\Phi^{(N)}
-
\overline{(1/\gamma)}_k
\underline\mu_k\underline\sigma_k
\Bigr]
\\
&\qquad\geq
(\underline\sigma_k+\overline\sigma_k)
\bigl[
(1-\bar\theta_k)\overline\nu_k^2
+\underline\sigma_k^2
\bigr]\underline\mu_k.
\end{aligned}
\]
This is the lower-endpoint inequality in the same assumption.
The assumptions impose the corresponding inequality
simultaneously for every \(i\in\mathcal I_k\).
Thus \(\sigma_k^*\), together with
\(\underline\mu_k\) and \(\overline\nu_k\), minimizes each
investor's Hamiltonian when evaluated at
the value function $v^i(t,x,y;\boldsymbol\vartheta^*)$.

\noindent
\textit{Step 3}.
Fix an investor $i$ and set $k=\iota_i$, so that
$i\in\mathcal I_k$. We verify that
\[
v^i(t,x,y;\boldsymbol\vartheta^*)
=
v^i(t,x,y;\pi^*,c^*,\boldsymbol\vartheta^{-k,*}).
\]
The fixed-market
HJB equation \eqref{eq:HJB-fixed}, evaluated at the Nash equilibrium
strategies under $\boldsymbol\vartheta^*$, gives
\begin{align*}
0={}&
v_t^i
-F_\mu^i\underline\mu_k
-\frac12F_\nu^i\overline\nu_k^{\,2}
-\left(
\frac12F_\nu^i(\sigma_k^*)^2
-F_\sigma^i\sigma_k^*
\right)
\\
&+
v_y^i\sum_{\ell\ne k}
\underline\mu_\ell\bar\pi_t^{\ell,*}
+
\frac12v_{yy}^i
\left[
\left(
\sum_{\ell\ne k}
\sigma_\ell^*\bar\pi_t^{\ell,*}
\right)^2
+
\sum_{\ell\ne k}
\left(
\overline\nu_\ell\bar\pi_t^{\ell,*}
\right)^2
\right]
\\
&-
\kappa\left[
c_t^{i,*}v_x^i
+\widehat c_t^{\,i,*}v_y^i
+\lambda_i
\exp\!\left(
-\delta_it
-\gamma_i\left(1-\frac{\theta_i}{N}\right)c_t^{i,*}
+\gamma_i\theta_i\widehat c_t^{\,i,*}
\right)
\right].
\end{align*}
Here all derivatives are those of
\(v^i(t,x,y;\boldsymbol\vartheta^*)\). By the pointwise
comparisons established above,
\[
\begin{aligned}
F_\mu^i\mu
&\leq F_\mu^i\underline\mu_k,\\
\frac12F_\nu^i\nu^2
&\leq\frac12F_\nu^i\overline\nu_k^{\,2},\\
\frac12F_\nu^i\sigma^2-F_\sigma^i\sigma
&\leq
\frac12F_\nu^i(\sigma_k^*)^2
-F_\sigma^i\sigma_k^*
\end{aligned}
\]
for every admissible \(\boldsymbol\vartheta=(\mu,\nu,\sigma)\) of asset \(k\). Consequently, the preceding equation is
also the HJB equation 
\eqref{eq:individual-robust-HJB}: its three suprema are attained
at \((\underline\mu_k,\overline\nu_k,\sigma_k^*)\).

Let \(\vartheta^k=(\mu^k,\nu^k,\sigma^k)\) be any admissible,
possibly progressively measurable, parameter process. Evolve
\((X^i,\widehat X^{\,i})\) under
\((\vartheta^k,\boldsymbol\vartheta^{-k,*})\), keeping the
strategies \((\pi^*,c^*)\) fixed. It\^o's formula
shows that the drift of
\[
v^i(s,X_s^i,\widehat X_s^{\,i};
       \boldsymbol\vartheta^*)
-
\kappa\lambda_i\int_t^s
\exp\!\left(
-\delta_ir
-\gamma_i\left(1-\frac{\theta_i}{N}\right)c_r^{i,*}
+\gamma_i\theta_i\widehat c_r^{\,i,*}
\right)dr
\]
is
\[
\begin{aligned}
&
F_\mu^i(\underline\mu_k-\mu_s^k)
+
\frac12F_\nu^i
\bigl(\overline\nu_k^{\,2}-(\nu_s^k)^2\bigr)
\\
&\quad+
\left[
\frac12F_\nu^i(\sigma_k^*)^2
-F_\sigma^i\sigma_k^*
\right]
-
\left[
\frac12F_\nu^i(\sigma_s^k)^2
-F_\sigma^i\sigma_s^k
\right]
\geq0.
\end{aligned}
\]
By taking conditional expectation with respect to 
$(X_t^i,\widehat X_t^{\,i})=(x,y)$, we obtain, for every
admissible $\vartheta^k$,
\[
\begin{aligned}
v^i(t,x,y;\boldsymbol\vartheta^*)
\leq
\mathbb E\Bigg[
&
e^{-\delta_iT}
U_i\!\left(
\left(1-\frac{\theta_i}{N}\right)X_T^i,
\widehat X_T^{\,i};
\gamma_i,\theta_i
\right)\\
&+
\left.\kappa\lambda_i\int_t^T e^{-\delta_is}
U_i\!\left(
\left(1-\frac{\theta_i}{N}\right)c_s^{i,*},
\widehat c_s^{\,i,*};
\gamma_i,\theta_i
\right)ds\right|(X_t^i,\widehat X_t^{\,i})=(x,y)
\Bigg].
\end{aligned}
\]
and equality holds for
$\vartheta^k=\vartheta^{k,*}$.
Taking the infimum over $\vartheta^k$ gives
\[
v^i(t,x,y;\pi^*,c^*,
     \boldsymbol\vartheta^{-k,*})
=
v^i(t,x,y;\boldsymbol\vartheta^*).
\]
In particular, at the initial state,
\[
\mathcal J_i(\pi^*,c^*;\boldsymbol\vartheta^*)
\leq
\mathcal J_i\bigl(
\pi^*,c^*;
\vartheta^k,\boldsymbol\vartheta^{-k,*}
\bigr).
\]
On the other hand, 
Theorem~\ref{thm:fixed-market-equilibrium} gives
\[
\mathcal J_i\bigl(
\pi^i,c^i,\pi^{-i,*},c^{-i,*};
\boldsymbol\vartheta^*
\bigr)
\leq
\mathcal J_i(\pi^*,c^*;\boldsymbol\vartheta^*)
\]
for every admissible unilateral deviation
$(\pi^i,c^i)$. 
Hence $\bigl((\pi^*,c^*),\boldsymbol\vartheta^*\bigr)$
is a robust Nash equilibrium.
\end{proof}




\subsection{Examples and economic implications}

We conclude this section with two examples illustrating the economic
implications of the robust Nash equilibrium. The first example shows that, in a
single-asset economy, competition affects investors' equilibrium portfolios
but does not alter their perception of model uncertainty. The second example
demonstrates that this property no longer holds in a multi-asset economy,
where sufficiently strong competition may change the worst-case model
itself.


\begin{example}[Single common stock]
\label{ex:single-stock} Suppose that all investors trade the same risky asset, so that $K=1$.
Assume that the asset has a strictly positive drift lower bound
$\underline\mu>0$, no asset-specific volatility
$\underline\nu=\overline\nu=0$, and uncertain common volatility satisfying
$0<\underline\sigma<\overline\sigma$. Let $\lambda_i=1$ and
$\delta_i=0$ for every investor, and assume that
$0\leq\bar\theta<1$.

Since there is only one risky asset, the aggregate competition weight is
$\bar\theta=N^{-1}\sum_{j=1}^N\theta_j$, and the relevant uncertainty is
the common volatility of this single asset. In this setting,
Assumption~\ref{ass:deterministic-worst-case} is automatically satisfied.
The compatible worst-case parameter profile is therefore
\(
\mu_t^*=\underline\mu,
\
\sigma_t^*=\overline\sigma.
\)

The robust Nash equilibrium is given by
\[
\pi_t^{i,*}
=
\left(
\frac{\theta_i\overline{(1/\gamma)}}{1-\bar\theta}
+
\frac{1}{\gamma_i}
\right)
\frac{\underline\mu}
{\overline\sigma^2\Upsilon_t},
\]
and
\[
c_t^{i,*}
=
\kappa
\left[
X_t^{i,*}\Upsilon_t
-
\left(
\frac{\theta_i}{1-\bar\theta}
\overline{(1/\gamma)}
+
\frac{1}{\gamma_i}
\right)
\frac{\underline\mu}{4\overline\sigma^2}
\left(
\Upsilon_t-\Upsilon_t^{-1}
\right)
\right].
\]

The investment formula separates two economic motives. The term
$1/\gamma_i$ is investor $i$'s direct risk return demand and is present
even when $\theta_i=0$. The term
$\theta_i\overline{(1/\gamma)}/(1-\bar\theta)$ is generated by relative-performance concerns. It is increasing in investor $i$'s competition
weight $\theta_i$ and in the aggregate risk tolerance of the population.
Hence, holding the market coefficients fixed, stronger relative-performance concerns increase the investor's equilibrium demand for the
risky asset under the present parameter restrictions.

The important feature of this single asset economy is that competition
does not alter the worst-case parameter selection. Regardless of the competition weights,
the largest admissible common volatility $\overline\sigma$ is selected.

\end{example}


\medskip

\begin{example}[Competition changes the worst-case model]
\label{ex:competition-model-uncertainty}
Consider two stocks, indexed by $k=1,2$. Their uncertainty sets satisfy
$0<\underline\mu_k\leq\overline\mu_k<\infty$,
$0\leq\underline\sigma_k<\overline\sigma_k<\infty$, and
$0<\underline\nu_k\leq\overline\nu_k<\infty$. For simplicity, suppose
that the two stocks have identical uncertainty bounds. Each stock
is traded by $N/2$ investors, so each trading group has population proportion $1/2$.

All investors trading stock $1$ have the same risk-aversion parameter
$\gamma_1$ and the same competition weight $\theta$. Investors trading
stock $2$ have the same risk-aversion parameter $\gamma_2$ and set
their competition weight equal to zero. Thus, the first group is
concerned with relative performance, whereas the second group is not.

\medskip

\noindent
\textit{1. The benchmark group without relative-performance concerns.}
For investors trading stock $2$, the competition weight is zero.
Theorem~\ref{thm:deterministic-robust-equilibrium} therefore gives
$\mu_2^*=\underline\mu$, $\nu_2^*=\overline\nu$, and
$\sigma_2^*=\overline\sigma$. Their robust Nash equilibrium  investment is
\[
\pi_t^{2,*}
=
\frac{\underline\mu}
{\gamma_2\bigl(\overline\nu^2+\overline\sigma^2\bigr)\Upsilon_t}.
\]
Thus, for this group, the worst-case model is the usual robust Merton
model: the worst-case drift is the lowest admissible drift, the
asset-specific volatility is the highest admissible volatility, and the
common volatility is also the highest admissible volatility.
This group provides a benchmark against which the effect of relative
performance for investors trading stock $1$ can be isolated.

\medskip

\noindent
\textit{2. Weak relative-performance concerns.} For the investors trading stock $1$, define
\[
\overline C
:=
\frac{\overline\sigma+\underline\sigma}
{2\overline\sigma+
(\overline\sigma-\underline\sigma)
\overline\sigma^2/\overline\nu^2},
\]
and let
\[
\theta_{\mathrm{weak}}
:=
\left(1+\frac{\overline\sigma^2}{\overline\nu^2}\right)
\left(
1+\overline C\frac{\gamma_2}{\gamma_1}
-
\sqrt{
1-
\frac{2\overline C}
{1+\overline\sigma^2/\overline\nu^2}
\frac{\gamma_2}{\gamma_1}
+
\overline C^2
\left(\frac{\gamma_2}{\gamma_1}\right)^2
}
\right).
\]
If $\theta\leq\theta_{\mathrm{weak}}$, then
$\sigma_1^*=\overline\sigma$, and
\[
\pi_t^{1,*}
=
\left[
\frac{\theta\overline\sigma^2}
{2\gamma_2(\overline\nu^2+\overline\sigma^2)}
+
\frac1{\gamma_1}
\right]
\frac{2\underline\mu}
{(2-\theta)
(\overline\nu^2+\overline\sigma^2)\Upsilon_t}.
\]

The selected worst-case model is therefore the same as in the
no-competition benchmark. Nevertheless, competition changes the
equilibrium investment strategies. In particular, the first term inside the square
brackets is increasing in $\theta$ and represents the incentive to
respond to the performance benchmark. Under the displayed condition,
this additional relative-performance motive increases the equilibrium
allocation to stock $1$, while the worst-case volatility remains
$\overline\sigma$. 
\medskip

\noindent
\textit{3. Strong relative-performance concerns.} Now define
\[
\underline C
:=
\frac{
(\overline\sigma+\underline\sigma)\overline\nu^2
}{
\overline\sigma
\bigl[
2\overline\nu^2
-
\underline\sigma(\overline\sigma-\underline\sigma)
\bigr]
},
\]
and let
\[
\begin{aligned}
\theta_{\mathrm{strong}}
:={}&
\left(1+\frac{\underline\sigma^2}{\overline\nu^2}\right)
+
\underline C
\left(1+\frac{\overline\sigma^2}{\overline\nu^2}\right)
\frac{\gamma_2}{\gamma_1}
\\
&-
\sqrt{
\left(1+\frac{\underline\sigma^2}{\overline\nu^2}\right)^2
-
2\underline C
\left(1+\frac{\overline\sigma^2}{\overline\nu^2}\right)
\frac{\gamma_2}{\gamma_1}
+
\underline C^2
\left(1+\frac{\overline\sigma^2}{\overline\nu^2}\right)^2
\left(\frac{\gamma_2}{\gamma_1}\right)^2
}.
\end{aligned}
\]
If $\theta\geq\theta_{\mathrm{strong}}$, then
$\sigma_1^*=\underline\sigma$, and
\[
\pi_t^{1,*}
=
\left[
\frac{\theta\underline\sigma\overline\sigma}
{2\gamma_2(\overline\nu^2+\overline\sigma^2)}
+
\frac1{\gamma_1}
\right]
\frac{2\underline\mu}
{(2-\theta)
(\overline\nu^2+\underline\sigma^2)\Upsilon_t}.
\]

This regime has a different economic interpretation. When relative
performance is sufficiently important, volatility affects not only the
risk of the investor's own portfolio but also the volatility of the
benchmark against which their performance is evaluated. A high-volatility
environment is therefore not necessarily the most adverse environment
from the investor's relative-performance perspective. A low-volatility
environment may be worse because it can make it more difficult to
generate an advantage over the market average. Accordingly, sufficiently strong relative-performance concerns reverse
the worst-case volatility selection: the worst-case  model for the
investors trading stock $1$ becomes the low-volatility model
$\sigma_1^*=\underline\sigma$, rather than the high-volatility model
$\sigma_1^*=\overline\sigma$.

The ratio $\gamma_2/\gamma_1$ shows that this selection is affected by
the characteristics of the second trading group, even though its
investors do not themselves care about relative performance. Their
investment contributes to the aggregate benchmark exposure faced by the
first group. Hence, investors in one trading group can affect both the
equilibrium investment strategies and the worst-case parameters selected by investors in
another group.

The example therefore illustrates two separate effects of competition:
for weak relative-performance concerns, competition changes the
equilibrium portfolio while leaving the worst-case model unchanged; for
strong relative-performance concerns, competition changes both the
equilibrium portfolio and the endogenous selection of the worst-case
market model.
\end{example}



\section{Robust mean-field game and asymptotic equilibrium}

In the previous section, we established the explicit robust Nash equilibrium
for the finite $N$-player investment--consumption game and characterized the
corresponding compatible worst-case parameter profile. In this section, we study the
large-population limit as $N\to\infty$ and derive the corresponding robust
mean-field game (MFG). As in the classical mean-field game theory
\cite{CD2018,LZ2019}, the limiting equilibrium provides an asymptotic
approximation of the finite-player Nash equilibrium.

Throughout this section, the number of tradable assets $K$ is fixed as $N\to\infty$, and each
investor trades one asset. Assume that the group proportions
$p_k^{(N)}=|\mathcal I_k|/N$ converge to $p_k>0$ for every
$k=1,\ldots,K$, where $\sum_{k=1}^K p_k=1$.
Following \cite{LZ2019}, we assume that the investor characteristics are
described by a random vector
\(
(x,\gamma,\theta)
\in
\mathbb R
\times
[\underline\gamma,\overline\gamma]
\times
[\underline\theta,\overline\theta],
\)
where
\(
0<\underline\gamma<\overline\gamma<\infty,
\
0\le\underline\theta<\overline\theta\le1.
\)
The initial wealth, risk aversion and competition weight of the investors are
assumed to be independent copies of $(x,\gamma,\theta)$ and are independent
of the Brownian motion
$(B,W^1,\ldots,W^K)$.

The finite-player game is coupled through the empirical averages of the
wealth, consumption and investment processes. As $N\to\infty$, the influence
of any individual investor on these averages becomes negligible, allowing
them to be replaced by mean-field processes. 
We first fix candidate mean-field processes and solve the
representative investor's robust control problem. This includes
verifying that the proposed deterministic market coefficients of each asset
are simultaneously worst-case for all representative investors
trading that asset. We then impose the mean-field consistency
conditions, requiring the candidate processes to coincide with the
conditional population averages generated by the resulting
strategies. The construction therefore proceeds in the following
two steps. 

\medskip

\noindent
\textit{Step 1. Solve the representative investor's robust control problem.}

Fix candidate mean-field processes
\(
(m^1,\ldots,m^K,X^0,c^0)
\)
and regard them as exogenous. The representative investor then solves an
auxiliary robust control problem using the corresponding robust-control verification arguments as in the finite-player game. This determines the optimal
investment--consumption strategy together with the individual worst-case
parameter selection for the traded asset. Since all representative investors
trading the same asset must face identical market coefficients, the
individual worst-case parameter selection is further required to be compatible,
yielding a compatible worst-case parameter profile.

\medskip

\noindent
\textit{Step 2. Recover the mean field through the consistency conditions.}

Unlike the finite-player game, the mean-field processes are not known a
priori. They are determined by imposing the mean-field consistency
conditions, which require the mean-field processes to coincide
with the population averages generated by the representative investors'
Nash equilibrium strategies under the compatible worst-case parameter profile.
This fixed-point condition determines the robust mean-field equilibrium.

\begin{figure}[h]
\centering
\begin{tikzpicture}[
    >=Latex,
    box/.style={
        draw=black,
        rounded corners=2pt,
        align=center,
        text width=5.4cm,
        inner sep=5pt,
        fill=gray!7,
        font=\footnotesize
    },
    arrow/.style={
        ->,
        thick,
        shorten >=2pt,
        shorten <=2pt
    },
    returnarrow/.style={
        ->,
        thick,
        rounded corners=3pt,
        shorten >=2pt,
        shorten <=2pt
    }
]

\node[box] (response) {
    \textbf{Step 1. representative investor's response}\\[1mm]
    Fix $(m^1,\ldots,m^K,X^0,c^0)$. Robust control yields
    type-dependent optimal strategies and a compatible
    worst-case parameter profile $\boldsymbol\vartheta^*$.
};

\node[box, right=2cm of response] (consistency) {
    \textbf{Step 2. Mean-field consistency}\\[1mm]
    Form conditional population averages of the Step~1 strategies,
    integrating over $(x,\gamma,\theta)$ given $\mathcal F_t$.
    Set them equal to the candidate mean-field processes.
};

\draw[arrow] (response.east) -- (consistency.west);

\coordinate (returnright) at ($(consistency.south)+(0,-9mm)$);
\coordinate (returnleft) at ($(response.south)+(0,-9mm)$);
\draw[returnarrow]
    (consistency.south) -- (returnright) -- (returnleft) -- (response.south);

\node[
    font=\scriptsize,
    fill=white,
    inner sep=2pt
] at ($(returnleft)!0.5!(returnright)$) {fixed-point consistency};

\end{tikzpicture}
\caption{The loop depicts the fixed-point dependence between the candidate
mean field processese and the conditional population averages they generate.}
\label{fig:robust-mfg-construction}
\end{figure}
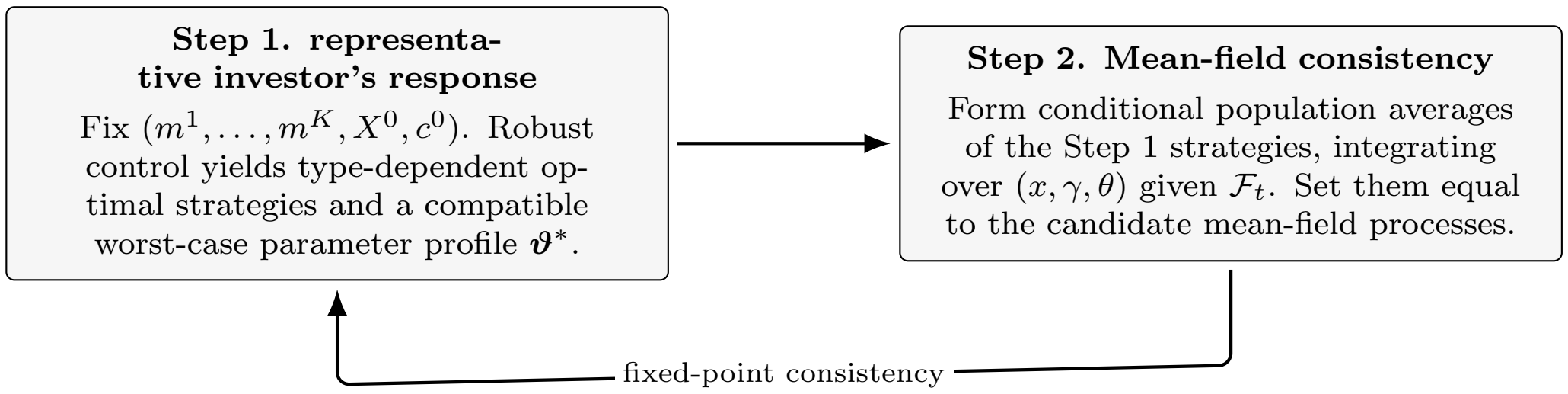

Figure~\ref{fig:robust-mfg-construction}
shows the resulting fixed-point structure.

\subsection{Auxiliary robust control problem and decentralized strategies}
\label{subsec:MF-auxiliary-control}

As discussed above, the empirical averages in the finite-player game are
replaced by their mean-field limits. For each asset $k=1,\ldots,K$, let
\(
m_t^k
\approx
\frac{1}{|\mathcal I_k|}
\sum_{j\in\mathcal I_k}\pi_t^j
\)
denote the average investment of the investors trading asset $k$.
Likewise, let $X^0_t\approx
\frac1N
\sum_{i=1}^N X_t^i$ and $c^0_t\approx
\frac1N
\sum_{i=1}^N c_t^i$ denote the limiting average wealth and
consumption processes, respectively. These processes will be determined later
through the mean-field consistency conditions.

Given candidate mean-field processes
\(
(m^1,\ldots,m^K,X^0,c^0),
\)
the limiting average wealth process satisfies
\begin{equation}
\label{eq:MF-average-wealth}
dX_t^0
={}
\sum_{k=1}^K
p_km_t^k
\left(
\mu_t^k\,dt
+
\sigma_t^k\,dB_t
+
\nu_t^k\,dW_t^k
\right)
-
\kappa c_t^0\,dt,
\qquad
X_0^0
={}
\mathbb E[x].
\end{equation}

For the representative investor $i$, define the relative wealth and
relative consumption processes by
\[
Z_t^i
=
X_t^i-\theta_iX_t^0,\qquad
C_t^i
=
c_t^i-\theta_ic_t^0.
\]
For fixed mean-field processes, the representative investor has a negligible
effect on the limiting population dynamics. Consequently, the interactions
with the remaining investors are completely represented by the 
mean-field processes together with the worst-case parameters of the
other trading groups.\\

\noindent
\textit{Representative investor's robust control problem.}
For the fixed mean-field processes
\(
(m^1,\ldots,m^K,X^0,c^0)
\)
and the worst-case parameter profile
\(
\vartheta_t^{k,*}
=
(\mu_t^{k,*},\nu_t^{k,*},\sigma_t^{k,*}),\)
\(
k\neq\iota_i,
\)
the representative investor chooses
\(
(\pi^i,C^i)\in\mathcal U_i
\)
while the market selects
\(
\vartheta^{\iota_i}
=
(\mu^{\iota_i},\nu^{\iota_i},\sigma^{\iota_i})
\in\mathcal U_{\iota_i}^{p}
\)
of the traded asset. The auxiliary robust value is
\begin{equation*}
v^i(0,z_i)
:=
\sup_{(\pi^i,C^i)\in\mathcal U_i}
\inf_{\vartheta^{\iota_i}\in\mathcal U_{\iota_i}^{p}}
J_i
\left(
\pi^i,C^i;
\vartheta^{\iota_i}
\right),
\end{equation*}
where the state process satisfies
\begin{equation*}
\left\{
\begin{aligned}
dZ_t^i
={}&
\Bigg[
\left(
\pi_t^i-\theta_ip_{\iota_i}m_t^{\iota_i}
\right)\mu_t^{\iota_i}
-
\kappa C_t^i
-
\theta_i
\sum_{k\neq\iota_i}
p_km_t^k\mu_t^{k,*}
\Bigg]dt
\\
&
+
\Bigg[
\left(
\pi_t^i-\theta_ip_{\iota_i}m_t^{\iota_i}
\right)\sigma_t^{\iota_i}
-
\theta_i
\sum_{k\neq\iota_i}
p_km_t^k\sigma_t^{k,*}
\Bigg]dB_t
\\
&
+
\left(
\pi_t^i-\theta_ip_{\iota_i}m_t^{\iota_i}
\right)
\nu_t^{\iota_i}\,dW_t^{\iota_i}
-
\theta_i
\sum_{k\neq\iota_i}
p_km_t^k\nu_t^{k,*}\,dW_t^k,
\\
Z_0^i
={}&
x_i-\theta_i\mathbb E[x],
\end{aligned}
\right.
\end{equation*}
and the payoff is
\begin{equation}\label{eq:MFG_payoff}
J_i
\left(
\pi^i,C^i;
\vartheta^{\iota_i}
\right)
=
-
\mathbb E
\left[
e^{-\gamma_iZ_T^i}
+
\kappa
\int_0^T
e^{-\gamma_iC_t^i}\,dt
\right].
\end{equation}
Note that, for simplicity, throughout this section we assume
\(
\lambda_i=1,
\
\delta_i=0
\) as compared with the $N$-player game payoff defined in (\ref{payoff_functional}).


\begin{theorem}
\label{thm:MF-RARC-solution}
Fix a representative investor $i$. Suppose that the candidate mean-field
investment strategies $m^1,\ldots,m^K$ are deterministic and nonnegative, and
that the worst-case parameters of the remaining trading groups are
prescribed by deterministic functions
\(
\vartheta_t^{k,*}
=
(\mu_t^{k,*},\nu_t^{k,*},\sigma_t^{k,*}),
\ k\neq\iota_i.
\)

Then the worst-case drift and asset-specific volatility of the asset traded by
the representative investor $i$ are
\(
\mu_t^{\iota_i,*}
=
\underline\mu_{\iota_i},
\
\nu_t^{\iota_i,*}
=
\overline\nu_{\iota_i}.
\)
Define the volatility switching function
\begin{equation*}
\begin{aligned}
\mathcal G_i^\sigma(\sigma,t)
:={}&
(\overline\sigma_{\iota_i}+\underline\sigma_{\iota_i})
\underline\mu_{\iota_i}-
\gamma_i\theta_i\Upsilon_t
\left[
2\overline\nu_{\iota_i}^{\,2}
+
\sigma
\left(
2\sigma-\overline\sigma_{\iota_i}-\underline\sigma_{\iota_i}
\right)
\right]
\sum_{k\neq\iota_i}
p_km_t^k\sigma_t^{k,*}.
\end{aligned}
\end{equation*}
The worst-case common volatility is given by
\begin{equation*}
\sigma_t^{\iota_i,*}
=
\begin{cases}
\overline\sigma_{\iota_i},
&
\mathcal G_i^\sigma(\overline\sigma_{\iota_i},t)\geq0,
\\[1mm]
\underline\sigma_{\iota_i},
&
\mathcal G_i^\sigma(\underline\sigma_{\iota_i},t)\leq0.
\end{cases}
\end{equation*}

Furthermore, the optimal
investment and relative consumption of the representative investor $i$ are
\begin{equation}
\label{eq:MF-optimal-pi-C}
\begin{aligned}
\pi_t^{i,\ast}
={}&
\frac{
\gamma_i\theta_i
\sigma_t^{\iota_i,*}
\left(
\sum_{k\neq\iota_i}
p_km_t^k\sigma_t^{k,*}
\right)\Upsilon_t
+
\underline\mu_{\iota_i}
}{
\gamma_i
\left[
\overline\nu_{\iota_i}^{\,2}
+
(\sigma_t^{\iota_i,*})^2
\right]\Upsilon_t
}
+
\theta_ip_{\iota_i}m_t^{\iota_i},
\\
C_t^{i,\ast}
={}&
-\frac{1}{\gamma_i}
\left[
\ln\Upsilon_t
+
f_t^i
-
\gamma_i\Upsilon_tZ_t^{i,\ast}
\right].
\end{aligned}
\end{equation}
Here
\(
\Upsilon_t=\frac{1}{1+\kappa(T-t)},
\)
$f^i$ is characterized explicitly by the linear ODE
\eqref{eq:MF-f-ODE} below, and $ Z^{i,\ast}$ denotes the corresponding optimal relative
wealth process.
\end{theorem}


\medskip



\begin{proof}
We divide the proof into four steps.

\noindent
\textit{Step 1.}
Fix an admissible strategy $(\pi^i,C^i)\in\mathcal U_i$. For the 
mean-field processes and the worst-case parameters of the remaining trading
groups, let $v^i(t,z;\pi^i,C^i,\boldsymbol\vartheta^{-\iota_i,*})$ denote the value function associated with the
worst-case selection of the coefficients of the asset traded by the representative investor $i$.
By the dynamic programming principle, $v^i$ satisfies
\begin{equation}
\label{eq:MF-fixed-control-HJB}
\begin{aligned}
0={}&
v_t^i
-
\sup_{\mu^{\iota_i}\in
[\underline\mu_{\iota_i},\overline\mu_{\iota_i}]}
\left\{
-
(\pi_t^i-\theta_ip_{\iota_i}m_t^{\iota_i})
\mu^{\iota_i}v_z^i
\right\}
\\
&
-
\sup_{\nu^{\iota_i}\in
[\underline\nu_{\iota_i},\overline\nu_{\iota_i}]}
\left\{
-\frac12
(\pi_t^i-\theta_ip_{\iota_i}m_t^{\iota_i})^2
(\nu^{\iota_i})^2v_{zz}^i
\right\}
\\
&
-
\sup_{\sigma^{\iota_i}\in
[\underline\sigma_{\iota_i},\overline\sigma_{\iota_i}]}
\Bigg\{
-\frac12
\left[
(\pi_t^i-\theta_ip_{\iota_i}m_t^{\iota_i})
\sigma^{\iota_i}
-
\theta_i
\sum_{k\neq\iota_i}
p_km_t^k\sigma_t^{k,*}
\right]^2
v_{zz}^i
\Bigg\}
\\
&
+
\frac{\theta_i^2}{2}
\sum_{k\neq\iota_i}
(p_km_t^k\nu_t^{k,*})^2v_{zz}^i
-
\left[
\kappa C_t^i
+
\theta_i
\sum_{k\neq\iota_i}
p_km_t^k\mu_t^{k,*}
\right]v_z^i
-
\kappa e^{-\gamma_iC_t^i},
\end{aligned}
\end{equation}
with terminal condition $v^i(T,z;\pi^i,C^i,\boldsymbol\vartheta^{-\iota_i,*})=-e^{-\gamma_i z}$.

Suppose temporarily that $v_z^i>0$ and $v_{zz}^i<0$. The worst-case drift is therefore
\[
\mu_t^{\iota_i,*}
\in
\begin{cases}
\{\underline\mu_{\iota_i}\},
&
\pi_t^i-\theta_ip_{\iota_i}m_t^{\iota_i}>0,
\\
[\underline\mu_{\iota_i},\overline\mu_{\iota_i}],
&
\pi_t^i-\theta_ip_{\iota_i}m_t^{\iota_i}=0,
\\
\{\overline\mu_{\iota_i}\},
&
\pi_t^i-\theta_ip_{\iota_i}m_t^{\iota_i}<0.
\end{cases}
\]
Similarly,
\[
\nu_t^{\iota_i,*}
\in
\begin{cases}
\{\overline\nu_{\iota_i}\},
&
\pi_t^i-\theta_ip_{\iota_i}m_t^{\iota_i}\neq0,
\\
[\underline\nu_{\iota_i},\overline\nu_{\iota_i}],
&
\pi_t^i-\theta_ip_{\iota_i}m_t^{\iota_i}=0.
\end{cases}
\]

For the common volatility, the worst-case selection is equivalent to maximizing
\[
\left[
(\pi_t^i-\theta_ip_{\iota_i}m_t^{\iota_i})\sigma
-
\theta_i
\sum_{k\neq\iota_i}
p_km_t^k\sigma_t^{k,*}
\right]^2
\]
over $[\underline\sigma_{\iota_i},\overline\sigma_{\iota_i}]$.
Hence the worst-case common volatility is attained at an endpoint. In the
region
$\pi_t^i-\theta_ip_{\iota_i}m_t^{\iota_i}>0$, which will contain the
optimal strategy, we obtain
\[
\sigma_t^{\iota_i,*}
\in
\begin{cases}
\{\overline\sigma_{\iota_i}\},
&
\pi_t^i-\theta_ip_{\iota_i}m_t^{\iota_i}
>
\dfrac{
2\theta_i
\sum_{k\neq\iota_i}
p_km_t^k\sigma_t^{k,*}
}{
\overline\sigma_{\iota_i}+\underline\sigma_{\iota_i}
},
\\[4mm]
\{\underline\sigma_{\iota_i},\overline\sigma_{\iota_i}\},
&
\pi_t^i-\theta_ip_{\iota_i}m_t^{\iota_i}
=
\dfrac{
2\theta_i
\sum_{k\neq\iota_i}
p_km_t^k\sigma_t^{k,*}
}{
\overline\sigma_{\iota_i}+\underline\sigma_{\iota_i}
},
\\[4mm]
\{\underline\sigma_{\iota_i}\},
&
0<
\pi_t^i-\theta_ip_{\iota_i}m_t^{\iota_i}
<
\dfrac{
2\theta_i
\sum_{k\neq\iota_i}
p_km_t^k\sigma_t^{k,*}
}{
\overline\sigma_{\iota_i}+\underline\sigma_{\iota_i}
}.
\end{cases}
\]


\medskip

\noindent
\textit{Step 2.}
Fix the candidate worst-case parameter profile
\(
\vartheta_t^{\iota_i,*}
=
(\mu_t^{\iota_i,*},\nu_t^{\iota_i,*},\sigma_t^{\iota_i,*})
\) as deterministic functions. Write $v^i(t,z;\boldsymbol\vartheta^*)$ for the value function of the fixed-market control problem.
The representative investor's HJB equation becomes
\begin{equation}
\label{eq:MF-control-HJB}
\begin{aligned}
0={}&
v_t^i
+
\sup_{\pi}
\Bigg\{
\frac12
\left(
\pi\sigma_t^{\iota_i,*}
-
\theta_i
\sum_{k=1}^{K}
p_km_t^k\sigma_t^{k,*}
\right)^2v_{zz}^i
\\
&\qquad
+
\frac12
\left(
\pi-\theta_ip_{\iota_i}m_t^{\iota_i}
\right)^2
(\nu_t^{\iota_i,*})^2v_{zz}^i
+
\left(
\pi-\theta_ip_{\iota_i}m_t^{\iota_i}
\right)
\mu_t^{\iota_i,*}v_z^i
\Bigg\}
\\
&
+
\frac{\theta_i^2}{2}
\sum_{k\neq\iota_i}
\left(
p_km_t^k\nu_t^{k,*}
\right)^2v_{zz}^i
-
\theta_i
\sum_{k\neq\iota_i}
p_km_t^k\mu_t^{k,*}v_z^i
+
\kappa
\sup_C
\left\{
-Cv_z^i-e^{-\gamma_iC}
\right\}.
\end{aligned}
\end{equation}

Assume temporarily that $v_z^i>0$ and $v_{zz}^i<0$. The first-order conditions yield
\begin{equation*}
\begin{aligned}
\pi_t^{i,\ast}
={}&
\frac{
\theta_i\sigma_t^{\iota_i,*}
\sum_{k\neq\iota_i}
p_km_t^k\sigma_t^{k,*}
}{
(\nu_t^{\iota_i,*})^2
+
(\sigma_t^{\iota_i,*})^2
}
+
\theta_ip_{\iota_i}m_t^{\iota_i}
-
\frac{
\mu_t^{\iota_i,*}v_z^i
}{
\left[
(\nu_t^{\iota_i,*})^2
+
(\sigma_t^{\iota_i,*})^2
\right]v_{zz}^i
},
\\
C_t^{i,\ast}
={}&
-\frac{1}{\gamma_i}
\ln\left(\frac{v_z^i}{\gamma_i}\right).
\end{aligned}
\end{equation*}

Motivated by the exponential utility, we seek a solution of the form
\[
v^i(t,z;\boldsymbol\vartheta^*)
=
-\exp\left(
f_t^i-\gamma_i\Upsilon_tz
\right),
\qquad
\Upsilon_t=\frac{1}{1+\kappa(T-t)}.
\]
Substitution into \eqref{eq:MF-control-HJB} shows that $f^i$
satisfies
\begin{equation}
\label{eq:MF-f-ODE}
\dot{f}_t^i
-
\kappa\Upsilon_tf_t^i
+
g_t^i
=
0,
\qquad
f_T^i=0,
\end{equation}
where
\begin{equation}\label{g_MFG}
\begin{aligned}
g_t^i
={}&
-
\frac{
\left[
\gamma_i\theta_i\sigma_t^{\iota_i,*}
\left(
\sum_{k\neq\iota_i}
p_km_t^k\sigma_t^{k,*}
\right)\Upsilon_t
+
\mu_t^{\iota_i,*}
\right]^2
}{
2
\left[
(\nu_t^{\iota_i,*})^2
+
(\sigma_t^{\iota_i,*})^2
\right]
}
+
\gamma_i\theta_i\Upsilon_t
\sum_{k\neq\iota_i}
p_km_t^k\mu_t^{k,*}
\\
&
+
\frac{\gamma_i^2\theta_i^2\Upsilon_t^2}{2}
\left[
\left(
\sum_{k\neq\iota_i}
p_km_t^k\sigma_t^{k,*}
\right)^2
+
\sum_{k\neq\iota_i}
\left(
p_km_t^k\nu_t^{k,*}
\right)^2
\right]
+
\kappa\Upsilon_t
\left(
1-\ln\Upsilon_t
\right).
\end{aligned}
\end{equation}
The unique solution of \eqref{eq:MF-f-ODE} is
\[
f_t^i
=
\Upsilon_t
\int_t^T
\frac{g_s^i}{\Upsilon_s}\,ds.
\]
Moreover,
\(
v_z^i
=
\gamma_i\Upsilon_t
e^{f_t^i-\gamma_i\Upsilon_tz}>0,
\
v_{zz}^i
=
-\gamma_i^2\Upsilon_t^2
e^{f_t^i-\gamma_i\Upsilon_tz}<0,
\)
verifying the sign conditions used above. Since
\(
-v_z^i/v_{zz}^i
=
1/(\gamma_i\Upsilon_t),
\)
the optimal controls are
\begin{equation}
\label{eq:MF-optimal-controls-general}
\begin{aligned}
\pi_t^{i,\ast}
={}&
\frac{
\gamma_i\theta_i\sigma_t^{\iota_i,*}
\left(
\sum_{k\neq\iota_i}
p_km_t^k\sigma_t^{k,*}
\right)\Upsilon_t
+
\mu_t^{\iota_i,*}
}{
\gamma_i
\left[
(\nu_t^{\iota_i,*})^2
+
(\sigma_t^{\iota_i,*})^2
\right]\Upsilon_t
}
+
\theta_ip_{\iota_i}m_t^{\iota_i},
\\
C_t^{i,\ast}
={}&
-\frac{1}{\gamma_i}
\left[
\ln\Upsilon_t
+
f_t^i
-
\gamma_i\Upsilon_tZ_t^{i,\ast}
\right].
\end{aligned}
\end{equation}

\medskip

\noindent
\textit{Step 3.}
From \eqref{eq:MF-optimal-controls-general}, the optimal investment strategy relative to
the mean field process is
\[
\pi_t^{i,\ast}-\theta_ip_{\iota_i}m_t^{\iota_i}
=
\frac{
\gamma_i\theta_i\sigma_t^{\iota_i,*}
\left(
\sum_{k\neq\iota_i}p_km_t^k\sigma_t^{k,*}
\right)\Upsilon_t
+
\mu_t^{\iota_i,*}
}{
\gamma_i
\left[
(\nu_t^{\iota_i,*})^2
+
(\sigma_t^{\iota_i,*})^2
\right]\Upsilon_t
}.
\]
Under the assumptions
$\underline\mu_{\iota_i}>0$, $\theta_i\geq0$,
$m_t^k\geq0$, and $\sigma_t^{k,*}\geq0$, the right-hand side is strictly
positive.
The worst-case characterization obtained in Step~1 therefore gives
\(
\mu_t^{\iota_i,*}
=
\underline\mu_{\iota_i},
\
\nu_t^{\iota_i,*}
=
\overline\nu_{\iota_i}.
\)
Substituting these values into
\eqref{eq:MF-optimal-controls-general} yields the optimal controls stated in
\eqref{eq:MF-optimal-pi-C}.

It remains to determine the worst-case common volatility. By
Step~1, the upper endpoint
$\overline\sigma_{\iota_i}$ is selected if
\[
\pi_t^{i,\ast}-\theta_ip_{\iota_i}m_t^{\iota_i}
\geq
\frac{
2\theta_i\sum_{k\neq\iota_i}
p_km_t^k\sigma_t^{k,*}
}{
\overline\sigma_{\iota_i}+\underline\sigma_{\iota_i}
}.
\]
Substituting the optimal investment strategy with
$\sigma_t^{\iota_i,*}=\overline\sigma_{\iota_i}$ and rearranging gives
\(
\mathcal G_i^\sigma
(\overline\sigma_{\iota_i},t)
\geq0.
\)
Similarly, the lower endpoint $\underline\sigma_{\iota_i}$ is selected if
\[
0<
\pi_t^{i,\ast}-\theta_ip_{\iota_i}m_t^{\iota_i}
\leq
\frac{
2\theta_i\sum_{k\neq\iota_i}
p_km_t^k\sigma_t^{k,*}
}{
\overline\sigma_{\iota_i}+\underline\sigma_{\iota_i}
}.
\]
Substituting the optimal investment strategy with
$\sigma_t^{\iota_i,*}=\underline\sigma_{\iota_i}$ and rearranging gives
\(\mathcal G_i^\sigma
(\underline\sigma_{\iota_i},t)
\leq0.
\) 

\noindent
\textit{Step 4}. 
Fix a representative investor $i$ assigned to asset
$k=\iota_i$. We verify that
\[
v^i(t,z;\boldsymbol\vartheta^*)
=
v^i(t,z;\pi^{i,*},C^{i,*},
       \boldsymbol\vartheta^{-k,*}).
\]
The fixed-market
HJB equation \eqref{eq:MF-control-HJB}, evaluated at the
optimal controls under $\boldsymbol\vartheta^*$, gives
\begin{align*}
0={}&v_t^i
+\bigl(\pi_t^{i,*}-\theta_i p_km_t^k\bigr)
  \underline\mu_k v_z^i
-\theta_i\sum_{\ell\ne k}p_\ell m_t^\ell
  \mu_t^{\ell,*}v_z^i
-\kappa\bigl[C_t^{i,*}v_z^i
             +e^{-\gamma_i C_t^{i,*}}\bigr]\\
&+\frac12v_{zz}^i
 \left[
 \bigl(\pi_t^{i,*}-\theta_i p_km_t^k\bigr)\sigma_t^{k,*}
 -\theta_i\sum_{\ell\ne k}p_\ell m_t^\ell
  \sigma_t^{\ell,*}
 \right]^2\\
&+\frac12v_{zz}^i
 \left[
 \bigl(\pi_t^{i,*}-\theta_i p_km_t^k\bigr)^2
 \overline\nu_k^{\,2}
 +\theta_i^2\sum_{\ell\ne k}
 (p_\ell m_t^\ell\nu_t^{\ell,*})^2
 \right].
\end{align*}
Here all derivatives are those of the fixed-market value function
$v^i(t,z;\boldsymbol\vartheta^*)$.
The comparisons in Step~3 show that, for every admissible
$(\mu,\nu,\sigma)$ of asset $k$,
\begin{align*}
\bigl(\pi_t^{i,*}-\theta_i p_km_t^k\bigr)\mu v_z^i
&\geq
\bigl(\pi_t^{i,*}-\theta_i p_km_t^k\bigr)
\underline\mu_k v_z^i,\\
\frac12\bigl(\pi_t^{i,*}-\theta_i p_km_t^k\bigr)^2
\nu^2v_{zz}^i
&\geq
\frac12\bigl(\pi_t^{i,*}-\theta_i p_km_t^k\bigr)^2
\overline\nu_k^{\,2}v_{zz}^i,\\
\frac12
\left[
\bigl(\pi_t^{i,*}-\theta_i p_km_t^k\bigr)\sigma
-\theta_i\sum_{\ell\ne k}p_\ell m_t^\ell
 \sigma_t^{\ell,*}
\right]^2v_{zz}^i
&\geq
\frac12
\left[
\bigl(\pi_t^{i,*}-\theta_i p_km_t^k\bigr)\sigma_t^{k,*}
-\theta_i\sum_{\ell\ne k}p_\ell m_t^\ell
 \sigma_t^{\ell,*}
\right]^2v_{zz}^i.
\end{align*}
Consequently, the preceding equation is
also the HJB equation 
\eqref{eq:MF-fixed-control-HJB}: its three suprema are attained
at \((\underline\mu_k,\overline\nu_k,\sigma_t^{k,*})\).

Let
$\vartheta^k=(\mu^k,\nu^k,\sigma^k)$ be any admissible,
possibly progressively measurable, parameter processes. Evolve $Z^i$
under under
$(\vartheta^k,\boldsymbol\vartheta^{-k,*})$, keeping the
strategies \((\pi^{i,*},C^{i,*})\) fixed. It\^o's formula shows that the drift of
\[
v^i(s,Z_s^i;\boldsymbol\vartheta^*)
-\kappa\int_t^s e^{-\gamma_i C_r^{i,*}}\,dr
\]
is
\begin{align*}
&\bigl(\pi_s^{i,*}-\theta_i p_km_s^k\bigr)
  (\mu_s^k-\underline\mu_k)v_z^i\\
&\quad+
\frac12\bigl(\pi_s^{i,*}-\theta_i p_km_s^k\bigr)^2
\bigl[(\nu_s^k)^2-\overline\nu_k^{\,2}\bigr]v_{zz}^i\\
&\quad+
\frac12v_{zz}^i
\left\{
\left[
\bigl(\pi_s^{i,*}-\theta_i p_km_s^k\bigr)\sigma_s^k
-\theta_i\sum_{\ell\ne k}p_\ell m_s^\ell
 \sigma_s^{\ell,*}
\right]^2
-
\left[
\bigl(\pi_s^{i,*}-\theta_i p_km_s^k\bigr)\sigma_s^{k,*}
-\theta_i\sum_{\ell\ne k}p_\ell m_s^\ell
 \sigma_s^{\ell,*}
\right]^2
\right\}
\geq0.
\end{align*}
By taking conditional expectation with respect to $Z_t^i=z$, we obtain that, for every admissible $\vartheta^k$,
\[
v^i(t,z;\boldsymbol\vartheta^*)
\leq
-\mathbb E\left[
\left.e^{-\gamma_i Z_T^i}
+
\kappa\int_t^T e^{-\gamma_i C_s^{i,*}}\,ds\right|Z_t^i=z
\right],
\]
and equality holds when $\vartheta^k=\vartheta^{k,*}$.
Taking the infimum over $\vartheta^k$ gives
\[
v^i(t,z;\pi^{i,*},C^{i,*},
       \boldsymbol\vartheta^{-k,*})
=
v^i(t,z;\boldsymbol\vartheta^*).
\]
In particular, at the initial state,  
\[
J_i(\pi^{i,*},C^{i,*};\boldsymbol\vartheta^{*})
\leq
J_i(\pi^{i,*},C^{i,*};\vartheta^k,\boldsymbol\vartheta^{-k,*}).
\]
On the other hand, verification of the fixed-market HJB equation
\eqref{eq:MF-control-HJB} in step 2 gives, for every admissible
unilateral deviation $(\pi^i,C^i)$,
\[
J_i(\pi^i,C^i;\boldsymbol\vartheta^{*})
\leq
J_i(\pi^{i,*},C^{i,*};\boldsymbol\vartheta^{*}).
\]
This completes the proof.
\end{proof}

\medskip




Theorem~\ref{thm:MF-RARC-solution} characterizes the optimal strategy and
individual worst-case parameters of a representative investor, conditional on
the worst-case parameters of the remaining trading groups. As in the
finite-player game, the individual selection must be compatible within each
trading group, since all investors trading the same asset face the same market
coefficients. We now give sufficient conditions that define a deterministic and compatible worst-case parameter profile
for all $K$ assets.

\begin{assumption}[Deterministic and compatible worst-case parameter conditions]
\label{ass:MF-deterministic-worst-case}
Assume the conditions of
Theorem~\ref{thm:MF-RARC-solution}.
Suppose further that the asset set admits a partition
\[
\{1,\ldots,K\}
=
\mathcal M\cup\overline{\mathcal M},
\qquad
\mathcal M\cap\overline{\mathcal M}
=
\varnothing.
\]

Define
\[
\Phi_t
:=
\sum_{k\in\mathcal M}
p_km_t^k\overline\sigma_k
+
\sum_{k\in\overline{\mathcal M}}
p_km_t^k\underline\sigma_k .
\]

Assume that

\begin{itemize}

\item
for every $k\in\mathcal M$,
\[
\overline\gamma\,\overline\theta\leq 
\frac{
(\underline\sigma_k+\overline\sigma_k)\underline\mu_k
}{
\bigl[
2\overline\nu_k^2
+
\overline\sigma_k
(\overline\sigma_k-\underline\sigma_k)
\bigr]
\bigl(
\Phi_t-p_km_t^k\overline\sigma_k
\bigr)
\Upsilon_t
};
\]

\item
for every $k\in\overline{\mathcal M}$,
\[
2\overline\nu_k^2
>
(\overline\sigma_k-\underline\sigma_k)
\underline\sigma_k,
\]
and
\[
\underline\gamma\,\underline\theta\geq 
\frac{
(\underline\sigma_k+\overline\sigma_k)\underline\mu_k
}{
\bigl[
2\overline\nu_k^2
-
\underline\sigma_k
(\overline\sigma_k-\underline\sigma_k)
\bigr]
\bigl(
\Phi_t-p_km_t^k\underline\sigma_k
\bigr)
\Upsilon_t
}.
\]

\end{itemize}

\end{assumption}


\begin{theorem}
\label{thm:MF-deterministic-robust-equilibrium}
Suppose
Assumption~\ref{ass:MF-deterministic-worst-case}
holds.
Then the individual worst-case parameter selection is independent of the
state variables and coincide for all the representative investors trading the
same asset. More precisely,
\[
\mu_t^{k,*}
=
\underline\mu_k,
\qquad
\nu_t^{k,*}
=
\overline\nu_k,
\qquad
\sigma_t^{k,*}=\sigma_k^*
=
\begin{cases}
\overline\sigma_k,
&
k\in\mathcal M,
\\[1mm]
\underline\sigma_k,
&
k\in\overline{\mathcal M},
\end{cases}
\qquad
k=1,\ldots,K.
\]

Consequently, the representative investor's robust control problem admits the
explicit solution
\eqref{eq:MF-optimal-pi-C},
where
\[
\sum_{k\neq\iota_i}
p_km_t^k\sigma_t^{k,*}
=
\Phi_t-p_{\iota_i}m_t^{\iota_i}\sigma_{\iota_i}^{*}.
\]

\end{theorem}


\begin{proof}
Fix $k\in\mathcal M$. Under 
Assumption~\ref{ass:MF-deterministic-worst-case},
\(
\Phi_t-p_km_t^k\overline\sigma_k>0.
\)
Moreover,
\[
\frac{
(\overline\sigma_k+\underline\sigma_k)\underline\mu_k
}{
\left[
2\overline\nu_k^{\,2}
+
\overline\sigma_k(\overline\sigma_k-\underline\sigma_k)
\right]
\Upsilon_t
(\Phi_t-p_km_t^k\overline\sigma_k)
}
\ge
\overline\gamma\,\overline\theta.
\]
For any $i\in\mathcal I_k$, since
\(
\gamma_i\theta_i
\le
\overline\gamma\,\overline\theta,
\)
we have
\(
\mathcal G_i^\sigma(\overline\sigma_k,t)\ge0.
\)
Hence, by Theorem~\ref{thm:MF-RARC-solution}, the representative investor
trading asset $k$ selects
\(
\sigma_t^{k,*}
=
\overline\sigma_k.
\)

Next, fix $k\in\overline{\mathcal M}$. Under 
Assumption~\ref{ass:MF-deterministic-worst-case},
\(
\Phi_t-p_km_t^k\underline\sigma_k>0
\)
and
\(
2\overline\nu_k^{\,2}
>
\underline\sigma_k
(\overline\sigma_k-\underline\sigma_k).
\)
Moreover,
\[
\frac{
(\overline\sigma_k+\underline\sigma_k)\underline\mu_k
}{
\left[
2\overline\nu_k^{\,2}
-
\underline\sigma_k(\overline\sigma_k-\underline\sigma_k)
\right]
\Upsilon_t
(\Phi_t-p_km_t^k\underline\sigma_k)
}
\le
\underline\gamma\,\underline\theta.
\]
For any $i\in\mathcal I_k$, since
\(
\gamma_i\theta_i
\ge
\underline\gamma\,\underline\theta,
\)
we have
\(
\mathcal G_i^\sigma(\underline\sigma_k,t)\le0.
\)
Hence, by Theorem~\ref{thm:MF-RARC-solution}, the representative investor
trading asset $k$ selects
\(
\sigma_t^{k,*}
=
\underline\sigma_k.
\)
\end{proof}



\subsection{Mean-field consistency conditions}
\label{subsec:MF-consistency}

The preceding subsection characterizes the optimal strategy of a
representative investor for the mean-field processes
\(
(m^1,\ldots,m^K,X^0,c^0)
\)
and a compatible worst-case parameter profile. These processes have not yet
been determined. A robust mean-field equilibrium requires the prescribed
mean-field quantities to coincide with the population averages generated by
the representative investors' optimal strategies. This requirement is the
usual mean-field \textit{consistency condition}, also referred to as the \textit{Nash
certainty equivalence principle}.

\begin{definition}[Mean-field consistency]
\label{Def:consistency}
For each $k=1,\ldots,K$, let $i_k$ denote a representative
investor assigned to group $k$, so that $\iota_{i_k}=k$. Let
\(
(\pi^{i_k,*},c^{i_k,*},X^{i_k,*})
\)
be the optimal controls and wealth process obtained in the preceding
subsection for the candidate mean-field processes
$(m^1,\ldots,m^K,X^0,c^0)$ and the candidate parameter profile.
The candidate mean-field processes satisfy the
{mean-field consistency conditions} if
\[
m_t^k
=
\mathbb E\!\left[\pi_t^{i_k,*}\right],
\qquad k=1,\ldots,K,
\]
and
\[
X_t^0
=
\sum_{k=1}^K p_k
\mathbb E\!\left[
X_t^{i_k,*}\,\middle|\,\mathcal F_t
\right],
\qquad
c_t^0
=
\sum_{k=1}^K p_k
\mathbb E\!\left[
c_t^{i_k,*}\,\middle|\,\mathcal F_t
\right].
\]
The expectations are taken over the representative investor types, with conditional expectations taken relative to
the Brownian filtration $\mathbb F$.
\end{definition}



\begin{theorem}
\label{thm:MF-consistency}
Suppose that Assumption~\ref{ass:MF-deterministic-worst-case} holds and 
that $\mathbb E[\theta]<1$. Then the mean-field consistency
system in Definition \ref{Def:consistency} admits a unique solution.

Specifically, define 
\begin{equation}
\label{eq:MF-common-noise-coefficient}
\Phi^0
:=
\frac{
\displaystyle
\mathbb E
\!\left[\frac{1}{\gamma}\right]
\sum_{k=1}^K
\frac{
p_k\sigma_k^*\underline\mu_k
}{
\bigl(1-p_k\mathbb E[\theta]\bigr)
\overline\nu_k^{\,2}
+
(\sigma_k^*)^2
}
}{
\displaystyle
1-
\mathbb E[\theta]
\sum_{k=1}^K
\frac{
p_k(\sigma_k^*)^2
}{
\bigl(1-p_k\mathbb E[\theta]\bigr)
\overline\nu_k^{\,2}
+
(\sigma_k^*)^2
}
}.
\end{equation}
Then, for each $k=1,\ldots,K$,
\begin{equation}
\label{eq:MF-m-process}
m_t^k
=
\frac{m_k}{\Upsilon_t},
\qquad
m_k
:=
\frac{
\sigma_k^*\Phi^0\,\mathbb E[\theta]
+
\underline\mu_k
\mathbb E\!\left[\frac{1}{\gamma}\right]
}{
(1-p_k\mathbb E[\theta])\overline\nu_k^{\,2}
+
(\sigma_k^*)^2
}.
\end{equation}
The average consumption process is
\begin{equation}
\label{eq:MF-c0}
c_t^0
=
\Upsilon_tX_t^0
-
\frac{
\mathbb E
\!\left[
\frac{\ln\Upsilon_t+\bar f_t}{\gamma}
\right]
}{
1-\mathbb E[\theta]
},
\end{equation}
and the average wealth process is
\begin{equation}
\label{eq:MF-X0}
\begin{aligned}
X_t^0
=
\frac{1}{\Upsilon_t}
\Bigg[
&
\Upsilon_0\mathbb E[x]
+
\int_0^t
\Bigg(
\sum_{k=1}^Kp_km_k\underline\mu_k
+
\frac{
\kappa\Upsilon_s
}{
1-\mathbb E[\theta]
}
\mathbb E
\!\left[
\frac{\ln\Upsilon_s+\bar f_s}{\gamma}
\right]
\Bigg)ds
\\
&
+
\sum_{k=1}^K
p_km_k\overline\nu_kW_t^k
+
\Phi^0B_t
\Bigg].
\end{aligned}
\end{equation}
Here $f_t$ denotes the type-dependent coefficient obtained from
$f_t^i$ by replacing $(x_i,\gamma_i,\theta_i)$ with a representative
investor type $(x,\gamma,\theta)$.

\end{theorem}


\medskip
\begin{proof}
We divide the proof into three steps.

\noindent
\textit{Step 1.}
For each trading group $k$, the investment consistency condition in Definition \ref{Def:consistency} is
\(
m_t^k
=
\mathbb E
\left[
\pi_t^{i_k,\ast}
\right].
\)
Since the type distribution is the same across trading groups,
substituting \eqref{eq:MF-optimal-pi-C} gives
\[
\begin{aligned}
m_t^k
=
\frac{
\sigma_k^*
\left(
\sum_{\ell\neq k}p_\ell m_t^\ell\sigma_\ell^*
\right)
\mathbb E[\theta]
}{
\overline\nu_k^{\,2}+(\sigma_k^*)^2
}
+
p_km_t^k\mathbb E[\theta]
+
\frac{
\underline\mu_k
\mathbb E\!\left[\frac{1}{\gamma}\right]
}{
\left[
\overline\nu_k^{\,2}+(\sigma_k^*)^2
\right]\Upsilon_t
}.
\end{aligned}
\]
Using
\(
\sum_{\ell\neq k}p_\ell m_t^\ell\sigma_\ell^*
=
\sum_{\ell=1}^Kp_\ell m_t^\ell\sigma_\ell^*
-
p_km_t^k\sigma_k^*,
\)
we obtain
\begin{equation}\label{eq:MF-investment-consistency}
m_t^k
=
\frac{
\sigma_k^*
\left(
\sum_{\ell=1}^Kp_\ell m_t^\ell\sigma_\ell^*
\right)
\Upsilon_t\mathbb E[\theta]
+
\underline\mu_k
\mathbb E\!\left[\frac{1}{\gamma}\right]
}{
\left[
(1-p_k\mathbb E[\theta])\overline\nu_k^{\,2}
+
(\sigma_k^*)^2
\right]\Upsilon_t
}.
\end{equation}

Multiplying this identity by $p_k\sigma_k^*$ and summing over
$k=1,\ldots,K$ yields
\[
\begin{aligned}
\sum_{k=1}^Kp_km_t^k\sigma_k^*
={}&
\mathbb E[\theta]
\left(
\sum_{k=1}^K
\frac{
p_k(\sigma_k^*)^2
}{
(1-p_k\mathbb E[\theta])\overline\nu_k^{\,2}
+
(\sigma_k^*)^2
}
\right)
\sum_{k=1}^Kp_km_t^k\sigma_k^*
\\
&+
\frac{
\mathbb E\!\left[\frac{1}{\gamma}\right]
}{
\Upsilon_t
}
\sum_{k=1}^K
\frac{
p_k\sigma_k^*\underline\mu_k
}{
(1-p_k\mathbb E[\theta])\overline\nu_k^{\,2}
+
(\sigma_k^*)^2
}.
\end{aligned}
\]
Moreover,
\(
\frac{
(\sigma_k^*)^2
}{
(1-p_k\mathbb E[\theta])\overline\nu_k^{\,2}
+
(\sigma_k^*)^2
}
\leq1,
\)
and hence
\[
\mathbb E[\theta]
\sum_{k=1}^K
\frac{
p_k(\sigma_k^*)^2
}{
(1-p_k\mathbb E[\theta])\overline\nu_k^{\,2}
+
(\sigma_k^*)^2
}
\leq
\mathbb E[\theta]
\sum_{k=1}^Kp_k
=
\mathbb E[\theta]
<
1.
\]
Therefore, the coefficient of
$\sum_{k=1}^Kp_km_t^k\sigma_k^*$ on the left-hand side is strictly positive,
and the equation has the unique solution
\(
\sum_{k=1}^Kp_km_t^k\sigma_k^*
=
\frac{\Phi^0}{\Upsilon_t},
\)
where $\Phi^0$ is defined by
\eqref{eq:MF-common-noise-coefficient}.
Substituting
this identity 
into \eqref{eq:MF-investment-consistency} gives
$m_t^k=m_k/\Upsilon_t$ for every $k=1,\ldots,K$,
where $m_k$ is defined in \eqref{eq:MF-m-process}.

\medskip

\noindent
\textit{Step 2.}
For a representative investor $i_k$ trading asset $k$, the optimal relative
consumption satisfies
\[
C_t^{i_k,\ast}
=
\Upsilon_t
\left(
X_t^{i_k,\ast}-\theta_{i_k} X_t^0
\right)
-
\frac{\ln\Upsilon_t+ f_t^{i_k}}{\gamma_{i_k}}.
\]
Since $c_t^{i_k,\ast}= C_t^{i_k,\ast}+\theta_{i_k} c_t^0$, we obtain
\[
c_t^{i_k,\ast}
=
\Upsilon_t
\left(
X_t^{i_k,\ast}-\theta_{i_k} X_t^0
\right)
-
\frac{\ln\Upsilon_t+f_t^{i_k}}{\gamma_{i_k}}
+
\theta_{i_k} c_t^0.
\]
The consumption consistency condition in
Definition~\ref{Def:consistency} therefore gives
\[
\begin{aligned}
c_t^0
={}&
\sum_{k=1}^Kp_k
\mathbb E
\left[
c_t^{i_k,\ast}
\,\middle|\,
\mathcal F_t
\right]
\\
={}&
\Upsilon_t
\sum_{k=1}^Kp_k
\mathbb E
\left[
X_t^{i_k,\ast}
\,\middle|\,
\mathcal F_t
\right]
-
\Upsilon_t\mathbb E[\theta]X_t^0
-
\mathbb E
\left[
\frac{\ln\Upsilon_t+ f_t}{\gamma}
\right]
+
\mathbb E[\theta]c_t^0.
\end{aligned}
\]
Using the wealth consistency condition
\(
X_t^0
=
\sum_{k=1}^Kp_k
\mathbb E
\left[
 X_t^{i_k,\ast}
\,\middle|\,
\mathcal F_t
\right]
\) in Definition \ref{Def:consistency}, 
we obtain
\[
\bigl(1-\mathbb E[\theta]\bigr)c_t^0
=
\bigl(1-\mathbb E[\theta]\bigr)\Upsilon_tX_t^0
-
\mathbb E
\left[
\frac{\ln\Upsilon_t+f_t}{\gamma}
\right],
\]
which proves \eqref{eq:MF-c0}.

\medskip

\noindent
\textit{Step 3. }
Under the compatible worst-case parameter profile,
\eqref{eq:MF-average-wealth} becomes
\[
dX_t^0
=
\sum_{k=1}^Kp_km_t^k
\left(
\underline\mu_k\,dt
+
\sigma_k^*\,dB_t
+
\overline\nu_k\,dW_t^k
\right)
-
\kappa c_t^0\,dt.
\]
Substituting $m_t^k=m_k/\Upsilon_t$
and \eqref{eq:MF-c0}, we obtain
\[
\begin{aligned}
dX_t^0
={}&
\Bigg[
-\kappa\Upsilon_tX_t^0
+
\frac{1}{\Upsilon_t}
\sum_{k=1}^Kp_km_k\underline\mu_k
+
\frac{\kappa}{
1-\mathbb E[\theta]
}
\mathbb E
\left[
\frac{\ln\Upsilon_t+f_t}{\gamma}
\right]
\Bigg]dt
\\
&+
\frac{1}{\Upsilon_t}
\sum_{k=1}^Kp_km_k\overline\nu_k\,dW_t^k
+
\frac{\Phi^0}{\Upsilon_t}\,dB_t,
\qquad
X_0^0=\mathbb E[x].
\end{aligned}
\]
Since $\dot\Upsilon_t=\kappa\Upsilon_t^2$, the product rule gives
\[
\begin{aligned}
d\!\left(\Upsilon_tX_t^0\right)
={}&
\Bigg[
\sum_{k=1}^Kp_km_k\underline\mu_k
+
\frac{\kappa\Upsilon_t}{
1-\mathbb E[\theta]
}
\mathbb E
\left[
\frac{\ln\Upsilon_t+f_t}{\gamma}
\right]
\Bigg]dt
\\
&+
\sum_{k=1}^Kp_km_k\overline\nu_k\,dW_t^k
+
\Phi^0\,dB_t.
\end{aligned}
\]
Integrating from $0$ to $t$ and using
$X_0^0=\mathbb E[x]$ yields
\eqref{eq:MF-X0}. 
\end{proof}

\medskip

For prescribed mean-field processes, recall that $\Phi_t$ defined in Assumption \ref{ass:MF-deterministic-worst-case} denotes the
aggregate exposure to the common noise. At the consistency
solution, $m_t^k=m_k/\Upsilon_t$, and therefore
\begin{equation}\label{eq:MFG_Phi_process}
\Phi_t
=
\sum_{k=1}^Kp_km_t^k\sigma_k^*
=
\frac{1}{\Upsilon_t}
\sum_{k=1}^Kp_km_k\sigma_k^*
=
\frac{\Phi^0}{\Upsilon_t}.
\end{equation}
Thus $\Phi^0$ is the constant scaled exposure  to the common noise determined by the
consistency conditions.

Substituting the consistency solution from
Theorem~\ref{thm:MF-consistency} into the representative investor's
strategy \eqref{eq:MF-optimal-pi-C} gives the decentralized robust
mean-field equilibrium
\begin{equation}
\label{eq:robust-mean-field-equilibrium}
\pi_t^{i,*}
=
\frac{\Pi_i^*}{\Upsilon_t},
\qquad
c_t^{i,*}
=
\Upsilon_t X_t^{i,*}+F_t^{i,c},
\end{equation}
where
\begin{equation}
\label{eq:robust-mean-field-equilibrium-pi}
\Pi_i^*
=
\frac{
\theta_i
\left(
\sigma_{\iota_i}^*\Phi^0
+
p_{\iota_i}m_{\iota_i}\overline\nu_{\iota_i}^{\,2}
\right)
+
\underline\mu_{\iota_i}/\gamma_i
}{
\left[
\overline\nu_{\iota_i}^{\,2}
+
(\sigma_{\iota_i}^*)^2
\right]
},
\end{equation}
and
\begin{equation}
\label{eq:robust-mean-field-equilibrium-F}
F_t^{i,c}
=
-
\frac{\ln\Upsilon_t+f_t^i}{\gamma_i}
-
\frac{\theta_i}{1-\mathbb E[\theta]}
\mathbb E\!\left[
\frac{\ln\Upsilon_t+f_t^i}{\gamma_i}
\right].
\end{equation}
The corresponding equilibrium wealth process is
\begin{equation*}
\begin{aligned}
X_t^{i,*}
=
\frac{1}{\Upsilon_t}
\Bigg[
&\Upsilon_0 x_i
+
\int_0^t
\left(
\Pi_i^*\underline\mu_{\iota_i}
-
\kappa\Upsilon_s F_s^{i,c}
\right)ds\\
&+
\Pi_i^*\overline\nu_{\iota_i}W_t^{\iota_i}
+
\Pi_i^*\sigma_{\iota_i}^*B_t
\Bigg].
\end{aligned}
\end{equation*}


\begin{example}[Single common stock]
\label{ex:MF-single-stock}
Suppose that all representative investors trade the same risky asset, i.e.,
$K=1$, and assume
\(
\underline\mu>0,\
\underline\nu=\overline\nu=0,\
0<\underline\sigma<\overline\sigma.
\)

In this case,
Assumption~\ref{ass:MF-deterministic-worst-case}
is automatically satisfied. Hence the compatible worst-case parameter profile
reduces to
\(
\mu_t^*=\underline\mu,
\
\sigma_t^*=\overline\sigma.
\)
Moreover, the mean-field consistency equations admit the explicit solution
\[
m_t
=
\frac{
\underline\mu\,
\mathbb E[1/\gamma]
}{
\overline\sigma^2
\bigl(1-\mathbb E[\theta]\bigr)
\Upsilon_t
}.
\]
Consequently, the robust mean-field equilibrium is given by
\[
\pi_t^{i,\ast}
=
\left(
\frac{1}{\gamma_i}
+
\frac{
\theta_i
\mathbb E[1/\gamma]
}{
1-\mathbb E[\theta]
}
\right)
\frac{\underline\mu}
{\overline\sigma^2\Upsilon_t},
\]
and
\[
c_t^{i,\ast}
=
\Upsilon_t X_t^{i,\ast}
-
\left(
\frac{1}{\gamma_i}
+
\frac{
\theta_i
\mathbb E[1/\gamma]
}{
1-\mathbb E[\theta]
}
\right)
\frac{\underline\mu^2}
{4\overline\sigma^2}
\left(
\Upsilon_t-\Upsilon_t^{-1}
\right).
\]

This example highlights the difference between the finite-player and
mean-field formulations. As in the finite-player game, relative-performance
concerns do not affect the representative investor's worst-case assessment of
model uncertainty: regardless of the distribution of the competition weights,
the worst-case volatility is the largest admissible value
$\overline\sigma$, as in the classical robust Merton problem.
The distinction lies in the equilibrium interaction. In the finite-player
game, the Nash equilibrium strategy depends on the empirical averages of the
investors trading the same asset, whereas in the mean-field limit these
quantities are replaced by deterministic mean-field processes. Consequently,
the effect of competition is summarized entirely through the population
moments $\mathbb E[1/\gamma]$ and
$\mathbb E[\theta]$. 
\end{example}

\subsection{Asymptotic properties and robust $\epsilon$-Nash equilibrium approximation}
\label{subsec:epsilon-Nash}

We analyze the asymptotic relationship between the robust mean-field equilibrium and its finite-player counterpart. First, we establish that the sequence of $N$-player Nash equilibrium strategies converges to the mean-field strategy as $N \to \infty$. Second, we prove that the mean-field strategy constitutes an asymptotic $\epsilon_N$-Nash equilibrium for the $N$-player game, with $\epsilon_N \to 0$ as $N \to \infty$.


We first show that the deterministic and compatible worst-case parameter
conditions in Assumption \ref{ass:deterministic-worst-case} converge, as $N\to\infty$, to the
corresponding conditions in Assumption \ref{ass:MF-deterministic-worst-case}.

\begin{lemma} 
\label{lem:limit-worst-case-conditions}
Suppose that $p_k^{(N)}=|\mathcal I_k|/N\to p_k>0$ for every
$k=1,\ldots,K$, and that the investor types $(\gamma_i,\theta_i)$ are
i.i.d. with values in
$[\underline\gamma,\overline\gamma]\times
[\underline\theta,\overline\theta]$. Assume also that
\(
(\underline\gamma,\underline\theta),
\;(\overline\gamma,\overline\theta)
\in \operatorname{supp}\mathcal L(\gamma,\theta).
\)
Then, for each fixed $k$, almost
surely with respect to the sampled investor types, 
\[
\bar\theta_k^{(N)}
:=\frac1N\sum_{i\in\mathcal I_k}\theta_i
\rightarrow p_k\mathbb E[\theta],
\qquad
\overline{(1/\gamma)}_k^{(N)}
:=\frac1N\sum_{i\in\mathcal I_k}\frac1{\gamma_i}
\rightarrow p_k\mathbb E\!\left[\frac1\gamma\right],
\]
and
\[
\max_{i\in\mathcal I_k}\gamma_i\theta_i
\rightarrow\overline\gamma\,\overline\theta,
\qquad
\min_{i\in\mathcal I_k}\gamma_i\theta_i
\rightarrow\underline\gamma\,\underline\theta.
\]
Furthermore, $\Phi^{(N)}\to\Phi^0$
as defined in \eqref{eq:MF-common-noise-coefficient}.
Consequently, the quantities on both sides of each group condition in
Assumption \ref{ass:deterministic-worst-case} converge to the corresponding quantities in
Assumption \ref{ass:MF-deterministic-worst-case}.
\end{lemma}

\medskip

\begin{proof}
The law of large
numbers gives, almost surely,
\[
\frac{1}{|\mathcal I_k|}
\sum_{i\in\mathcal I_k}\theta_i
\rightarrow
\mathbb E[\theta],
\qquad
\frac{1}{|\mathcal I_k|}
\sum_{i\in\mathcal I_k}\frac{1}{\gamma_i}
\rightarrow
\mathbb E\!\left[\frac{1}{\gamma}\right].
\]
Multiplying by $|\mathcal I_k|/N$ yields
\[
\bar\theta_k^{(N)}
\rightarrow p_k\mathbb E[\theta],
\qquad
\overline{(1/\gamma)}_k^{(N)}
\rightarrow
p_k\mathbb E\!\left[\frac{1}{\gamma}\right].
\]

We consider the sample maximum; the argument for the sample minimum is
similar. The support assumption implies that, for every
$\varepsilon>0$,
\(
\mathbb P\!\left(
\gamma\theta>
\overline\gamma\,\overline\theta-\varepsilon
\right)>0.
\)
By independence of the investor types,
\[
\begin{aligned}
&\mathbb P\!\left(
\max_{i\in\mathcal I_k}\gamma_i\theta_i
\leq
\overline\gamma\,\overline\theta-\varepsilon
\right)=
\left[
1-
\mathbb P\!\left(
\gamma\theta>
\overline\gamma\,\overline\theta-\varepsilon
\right)
\right]^{|\mathcal I_k|}.
\end{aligned}
\]
Since $|\mathcal I_k|/N\to p_k>0$, the right-hand side is
summable over $N$. The first Borel--Cantelli lemma therefore
implies that, almost surely, 
\(
\max_{i\in\mathcal I_k}\gamma_i\theta_i
>
\overline\gamma\,\overline\theta-\varepsilon.
\)
Together with the upper
bound on $\gamma_i\theta_i$, we obtain, almost surely,
\(
\max_{i\in\mathcal I_k}\gamma_i\theta_i
\rightarrow
\overline\gamma\,\overline\theta.
\)

Substituting the limits of the group averages into the definition
of $\Phi^{(N)}$ and using
\eqref{eq:MF-common-noise-coefficient} gives
\(
\Phi^{(N)}\to\Phi^0
\)
almost surely.
It remains to compare the group thresholds. We consider
$k\in\mathcal M$, for which $\sigma_k^*=\overline\sigma_k$.
At the mean-field consistency solution,
\eqref{eq:MF-m-process} and (\ref{eq:MFG_Phi_process}) give $m_t^k=m_k/\Upsilon_t$ and
$\Phi_t=\Phi^0/\Upsilon_t$. Hence
\[
\begin{aligned}
&\frac{
(1-p_k\mathbb E[\theta])
(\overline\nu_k^{\,2}+\overline\sigma_k^2)\Phi^0
-
p_k\mathbb E[1/\gamma]\,
\underline\mu_k\overline\sigma_k
}{
(1-p_k\mathbb E[\theta])\overline\nu_k^{\,2}
+\overline\sigma_k^2
}
=
\Upsilon_t
\bigl(\Phi_t-p_km_t^k\overline\sigma_k\bigr).
\end{aligned}
\]
Therefore, using the limits of $\bar\theta_k^{(N)}$,
$\overline{(1/\gamma)}_k^{(N)}$ and $\Phi^{(N)}$, the
right-hand side of the condition for $k\in\mathcal M$ in
Assumption~\ref{ass:deterministic-worst-case} converges to
\[
\frac{
(\underline\sigma_k+\overline\sigma_k)\underline\mu_k
}{
\bigl[
2\overline\nu_k^{\,2}
+\overline\sigma_k(\overline\sigma_k-\underline\sigma_k)
\bigr]
\bigl(\Phi_t-p_km_t^k\overline\sigma_k\bigr)\Upsilon_t
},
\]
which is the corresponding threshold in
Assumption~\ref{ass:MF-deterministic-worst-case}.
The same calculation with $\sigma_k^*=\underline\sigma_k$
gives the corresponding conclusion for
$k\in\overline{\mathcal M}$. 
\end{proof}


\medskip

In the following comparison, we distinguish the mean-field equilibrium
$(\pi^{i,*},c^{i,*},X^{i,*})$ from the finite-player equilibrium
$(\pi^{i,N,*},c^{i,N,*},X^{i,N,*})$. We also take $\kappa=1$ in (\ref{eq:MFG_payoff}) in the rest of this section. 

\begin{proposition}
\label{prop:MF-convergence}
Suppose the assumptions of
Theorems~\ref{thm:deterministic-robust-equilibrium} and
\ref{thm:MF-consistency} hold, and that
the resulting Nash equilibria select the same compatible worst-case
parameter profile. Couple investor $i$ in the two games using the same
initial wealth and Brownian motion. 
Define
\[
d_N
:=
\max_{1\leq i\leq N}|\Pi_i^{N,*}-\Pi_i^*|
+
\max_{1\leq i\leq N}\sup_{0\leq t\leq T}
|F_t^{i,N,c}-F_t^{i,c}|.
\]
Then, almost surely with respect to the sampled investor types,
$
d_N\rightarrow0.
$
In particular,
\begin{equation}
\label{eq:convergence-pi}
\max_{1\leq i\leq N}\sup_{0\leq t\leq T}
|\pi_t^{i,N,*}-\pi_t^{i,*}|
\rightarrow0.
\end{equation}
Moreover, almost surely with respect to the sampled investor
types, 
\begin{equation}
\label{eq:MF-lifted-state-bound}
\begin{aligned}
\max_{1\leq i\leq N}
\mathbb E\Bigg[
\sup_{0\leq t\leq T}
|X_t^{i,N,*}-X_t^{i,*}|^2
+
\int_0^T|c_t^{i,N,*}-c_t^{i,*}|^2dt
\,\Bigg|\,
(x_i,\gamma_i, \theta_i)_{i=1}^N
\Bigg]
\leq Cd_N^2,
\end{aligned}
\end{equation}
where $C$ is independent of $i$ and $N$. Consequently, for each
fixed $i$,
\begin{equation}
\label{eq:MF-convergence-wealth-consumption}
\begin{aligned}
\mathbb E\Bigg[
\sup_{0\leq t\leq T}
|X_t^{i,N,*}-X_t^{i,*}|^2
+
\int_0^T|c_t^{i,N,*}-c_t^{i,*}|^2dt
\Bigg]
\rightarrow0,
\end{aligned}
\end{equation}
where the expectation is taken over both the sampled investor
types and the Brownian motion.

\end{proposition}

\medskip

\begin{proof}
By Lemma~\ref{lem:limit-worst-case-conditions},
$\Phi^{(N)}\to\Phi^0$. Summing
\eqref{eq:equilibrium-investment} over investors in
$\mathcal I_k$ and dividing by $N$ gives
\[
\Upsilon_t\bar\pi_t^{k,N,*}
=
\frac{
\bar\theta_k^{(N)}\sigma_k^*\Phi^{(N)}
+
\overline{(1/\gamma)}_k^{(N)}\underline\mu_k
}{
(1-\bar\theta_k^{(N)})\overline\nu_k^{\,2}
+
(\sigma_k^*)^2
}.
\]
The group-average limits and \eqref{eq:MF-m-process} imply
\begin{equation}
\label{eq:group-investment-convergence}
\max_{1\leq k\leq K}\sup_{0\leq t\leq T}
\left|
\Upsilon_t\bar\pi_t^{k,N,*}-p_km_k
\right|
\rightarrow0.
\end{equation}

For $k=\iota_i$,  recall that 
\[
\pi_t^{i,N,*} \Upsilon_t=\Pi_i^{N,*}
=
\frac{
\underline\mu_k/\gamma_i
+
\theta_i
\left(
\sigma_k^*\Phi^{(N)}
+
\overline\nu_k^{\,2}
\Upsilon_t\bar\pi_t^{k,N,*}
\right)
}{
\overline\nu_k^{\,2}+(\sigma_k^*)^2
},
\]
whereas
\[
\pi_t^{i,*} \Upsilon_t=\Pi_i^*
=
\frac{
\underline\mu_k/\gamma_i
+
\theta_i
\left(
\sigma_k^*\Phi^0
+
\overline\nu_k^{\,2}p_km_k
\right)
}{
\overline\nu_k^{\,2}+(\sigma_k^*)^2
}.
\]
It follows that 
\(
\max_{1\leq i\leq N}|\Pi_i^{N,*}-\Pi_i^*|
\rightarrow0.
\)
Since $\Upsilon$ is bounded away from zero, this also proves
\eqref{eq:convergence-pi}.

For consumption, write
\(
c_t^{i,N,*}
=
\Upsilon_tX_t^{i,N,*}+F_t^{i,N,c},
\)
where
\begin{equation}
\label{eq:finite-consumption-offset-convergence}
F_t^{i,N,c}
=
-\frac{\ln\Upsilon_t+f_t^{i,N}}{\gamma_i}
-
\frac{\theta_i}{1-\bar\theta^{(N)}}
\frac1N\sum_{j=1}^N
\frac{\ln\Upsilon_t+f_t^{j,N}}{\gamma_j},
\qquad
\bar\theta^{(N)}
=
\frac1N\sum_{j=1}^N\theta_j.
\end{equation}
The finite-player and mean-field ODE drivers have the same form,
with $\Upsilon_t\bar\pi_t^{\ell,N,*}$ replaced by
$p_\ell m_\ell$. The bounded type ranges 
and \eqref{eq:group-investment-convergence} give
\[
\max_{1\leq j\leq N}\sup_{0\leq t\leq T}
|g_t^{j,N}-g_t^j|
\rightarrow0,
\]
with $g^{j,N}$ and $g^j$ given in (\ref{g_finite_player}) and (\ref{g_MFG}), respectively. 
The integral representations of
\eqref{eq:fixed-market-f-ODE} and \eqref{eq:MF-f-ODE}
therefore imply
\[
\max_{1\leq j\leq N}\sup_{0\leq t\leq T}
|f_t^{j,N}-f_t^j|
\rightarrow0.
\]
For the empirical average in
\eqref{eq:finite-consumption-offset-convergence},
\[
\begin{aligned}
&\left|
\frac1N\sum_{j=1}^N
\frac{\ln\Upsilon_t+f_t^{j,N}}{\gamma_j}
-
\mathbb E\!\left[
\frac{\ln\Upsilon_t+f_t}{\gamma}
\right]
\right|
\\
&\quad\leq
\frac1{\underline\gamma}
\max_{1\leq j\leq N}\sup_{0\leq s\leq T}
|f_s^{j,N}-f_s^j|
+
\left|
\frac1N\sum_{j=1}^N
\frac{\ln\Upsilon_t+f_t^j}{\gamma_j}
-
\mathbb E\!\left[
\frac{\ln\Upsilon_t+f_t}{\gamma}
\right]
\right|\rightarrow 0.
\end{aligned}
\]
Comparison with
\eqref{eq:robust-mean-field-equilibrium-F} gives
\begin{equation}
\label{eq:finite-MF-F-convergence}
\max_{1\leq i\leq N}\sup_{0\leq t\leq T}
|F_t^{i,N,c}-F_t^{i,c}|
\rightarrow0,
\end{equation}
so that $d_N\rightarrow 0$.

Under the common worst-case parameter profile, the difference of the two wealth processes 
give, for $k=\iota_i$,
\[
\begin{aligned}
X_t^{i,N,*}-X_t^{i,*}
=
\frac1{\Upsilon_t}\Bigg[
&(\Pi_i^{N,*}-\Pi_i^*)
\bigl(
\underline\mu_k t
+
\overline\nu_kW_t^k
+
\sigma_k^*B_t
\bigr)
\\
&-
\int_0^t\Upsilon_s
\bigl(F_s^{i,N,c}-F_s^{i,c}\bigr)ds
\Bigg].
\end{aligned}
\]
Conditional on the investor types, the coefficient differences
are deterministic. Doob's $L^2$ maximal inequality therefore
gives
\[
\max_{1\leq i\leq N}
\mathbb E\left[
\sup_{0\leq t\leq T}
|X_t^{i,N,*}-X_t^{i,*}|^2
\,\middle|\,
(x_i,\gamma_i, \theta_i)_{i=1}^N
\right]
\leq Cd_N^2.
\]
Moreover,
\[
c_t^{i,N,*}-c_t^{i,*}
=
\Upsilon_t(X_t^{i,N,*}-X_t^{i,*})
+
F_t^{i,N,c}-F_t^{i,c}.
\]
Hence, 
\eqref{eq:MF-lifted-state-bound} and (\ref{eq:MF-convergence-wealth-consumption}) follow. 
\end{proof}


\medskip





Finally, to construct an approximate equilibrium for the $N$-player game,
we apply the mean-field strategy to each investor
$i=1,\ldots,N$ and write the strategy as $(\pi^{i,*},c^{i,*})_{i=1}^N$.
We also write $(\pi^{-i,*},c^{-i,*})$ for the strategies of all investors
except $i$.

\begin{definition}[Asymptotic robust $\epsilon$-Nash equilibrium]
\label{def:asymptotic-epsilon-Nash}
The sequence of mean-field strategy profiles
$\{(\pi^{i,*},c^{i,*})_{i=1}^N\}_{N\geq1}$ is an
{asymptotic robust $\epsilon$-Nash equilibrium} if there
exist parameter profiles $\boldsymbol\vartheta^N$ and
$\epsilon_N\geq0$ with $\epsilon_N\to0$ such that the following conditions hold:

\begin{enumerate}
\item[(i)] The parameter profile is compatible with the
mean-field strategies applied to the finite-player game:
\[
\boldsymbol\vartheta^N
\in
\Gamma^N((\pi^{i,*},c^{i,*})_{i=1}^N).
\]

\item[(ii)] For every investor $i=1,\ldots,N$ and every
admissible unilateral deviation $(\pi^i,c^i)\in\mathcal U_i$,
\begin{equation}
\label{eq:epsilon-Nash-definition}
\begin{aligned}
&\mathcal J_i^N\bigl(
\pi^{i,*},c^{i,*},
\pi^{-i,*},c^{-i,*};
\boldsymbol\vartheta^N
\bigr)
\\
&\qquad\geq
\mathcal J_i^N\bigl(
\pi^i,c^i,
\pi^{-i,*},c^{-i,*};
\boldsymbol\vartheta^N
\bigr)
-\epsilon_N.
\end{aligned}
\end{equation}
\end{enumerate}
\end{definition}





\begin{proposition}
\label{prop:MF-epsilon-Nash}
Suppose that the assumptions of
Proposition~\ref{prop:MF-convergence} hold, and assume that
$|x|\leq M$ almost surely for some $M<\infty$.
Let $\boldsymbol\vartheta^*$ be the deterministic worst-case
parameter profile selected by the mean-field equilibrium. Suppose
further that, almost surely with respect to the sampled investor
types,
\[
\boldsymbol\vartheta^*
\in
\Gamma^N\bigl((\pi^{i,*},c^{i,*})_{i=1}^N\bigr).
\]
Then, almost surely with respect to the sampled investor types,
the sequence of mean-field strategy profiles
\(
\bigl\{(\pi^{i,*},c^{i,*})_{i=1}^N\bigr\}_{N\geq1}
\)
is an asymptotic robust $\epsilon$-Nash equilibrium.
\end{proposition}

\medskip

\begin{proof}
Throughout the proof, expectations in the finite-player payoffs
are taken over the Brownian motion, and all payoffs are
evaluated under $\boldsymbol\vartheta^*$. By
Proposition \ref{prop:MF-convergence}, $d_N\to0$.
In the following, we divide the proof into three steps. 

\textit{Step 1.}
For prescribed opponents' strategy profile $(\pi^{-i},c^{-i})$, define investor $i$'s
value function by
\[
\mathcal V_i^N(\pi^{-i},c^{-i})
:=
\sup_{(\pi^i,c^i)\in\mathcal U_i}
\mathcal J_i^N\bigl(
\pi^i,c^i,\pi^{-i},c^{-i};
\boldsymbol\vartheta^*
\bigr).
\]
We first compare the above value functions corresponding to two opponents' strategy profiles
$(\pi^{-i,N,*},c^{-i,N,*})$ and
$(\pi^{-i,*},c^{-i,*})$. For every group $\mathcal I_k$,
\[
\begin{aligned}
&\left|
\frac1N
\sum_{\substack{j\in\mathcal I_k\\j\neq i}}
\bigl(\pi_t^{j,N,*}-\pi_t^{j,*}\bigr)
\right|
\leq
\frac{1}{N\Upsilon_t}
\sum_{\substack{j\in\mathcal I_k\\j\neq i}}
|\Pi_j^{N,*}-\Pi_j^*|
\leq
\frac{d_N}{\Upsilon_t}
\leq Cd_N.
\end{aligned}
\]
Both opponents' strategy profiles have deterministic investment strategies
and affine consumption strategies with wealth coefficient
$\Upsilon_t$. Hence, Steps 1 and 2
in the proof of Theorem~\ref{thm:fixed-market-equilibrium} apply
to both opponents' strategy profiles. 
The investor types
\(
|x_i|\leq M,\;
\gamma_i\in[\underline\gamma,\overline\gamma],\;
\theta_i\in[\underline\theta,\overline\theta]
\),
the investment coefficients
\((\Pi_i^{N,*},\Pi_i^*)\), and the market coefficients
\(\boldsymbol\vartheta^*\)
are uniformly bounded. Moreover,
\(
\min_{1\leq k\leq K}
\bigl[\overline\nu_k^{\,2}+(\sigma_k^*)^2\bigr]>0.
\)
The expressions in \eqref{g_finite_player} are therefore
Lipschitz in 
\(
\widehat\pi_t^{\,i,\iota_i}
\)
and
\(
(\bar\pi_t^\ell)_{\ell\neq\iota_i}
\).
It follows that the drivers in
\eqref{eq:fixed-market-f-ODE} corresponding to
$(\pi^{-i,N,*},c^{-i,N,*})$ and
$(\pi^{-i,*},c^{-i,*})$
differ by at most $Cd_N$.

For either opponent's strategy profile, write
$f_t^{i,N}(\pi^{-i},c^{-i})$ for the coefficient in
investor $i$'s fixed-market value function. The two coefficients
\[
f_t^{i,N}(\pi^{-i,N,*},c^{-i,N,*})
\quad\text{and}\quad
f_t^{i,N}(\pi^{-i,*},c^{-i,*})
\]
solve the linear ODE \eqref{eq:fixed-market-f-ODE} with terminal value zero. Their integral representations therefore
give
\[
\begin{aligned}
\max_{1\leq i\leq N}\sup_{0\leq t\leq T}
\Big|
f_t^{i,N}(\pi^{-i,N,*},c^{-i,N,*})
-
f_t^{i,N}(\pi^{-i,*},c^{-i,*})
\Big|
\leq Cd_N.
\end{aligned}
\]

Both problems have the same initial state
$(x_i,\widehat x_i)$, where
\(
\widehat x_i=\frac1N\sum_{j\neq i}x_j.
\)
For either opponent's strategy profile, the value function is
\[
\begin{aligned}
\mathcal V_i^N(\pi^{-i},c^{-i})
=
-\exp\Bigg(
f_0^{i,N}(\pi^{-i},c^{-i})
-
\gamma_i\Upsilon_0
\left[
\left(1-\frac{\theta_i}{N}\right)x_i
-\theta_i\widehat x_i
\right]
\Bigg).
\end{aligned}
\]
Since $|x_i|,|\widehat x_i|\leq M$, the exponents in the
value functions
\(
\mathcal V_i^N(\pi^{-i,N,*},c^{-i,N,*})\) and
\(
\mathcal V_i^N(\pi^{-i,*},c^{-i,*})
\)
are uniformly bounded. Applying the mean-value theorem to their
exponential expressions gives
\begin{equation}
\label{eq:MF-best-response-value-stability}
\begin{aligned}
\max_{1\leq i\leq N}
\Big|
\mathcal V_i^N(\pi^{-i,*},c^{-i,*})
-
\mathcal V_i^N(\pi^{-i,N,*},c^{-i,N,*})
\Big|
\leq Cd_N.
\end{aligned}
\end{equation}

\textit{Step 2.}
We next compare the payoffs under the finite-player equilibrium
strategies
\(
(\pi^{j,N,*},c^{j,N,*})_{j=1}^N
\)
and the mean-field strategies
\(
(\pi^{j,*},c^{j,*})_{j=1}^N.
\)
Combining Jensen's inequality with
\eqref{eq:MF-lifted-state-bound} gives
\[
\begin{aligned}
\max_{1\leq i\leq N}
\mathbb E\Bigg|
X_T^{i,N,*}-X_T^{i,*}
-
\frac{\theta_i}{N}\sum_{j=1}^N
\bigl(X_T^{j,N,*}-X_T^{j,*}\bigr)
\Bigg|^2
\leq Cd_N^2,
\end{aligned}
\]
and
\[
\begin{aligned}
\max_{1\leq i\leq N}
\mathbb E\int_0^T\Bigg|
c_t^{i,N,*}-c_t^{i,*}
-
\frac{\theta_i}{N}\sum_{j=1}^N
\bigl(c_t^{j,N,*}-c_t^{j,*}\bigr)
\Bigg|^2dt
\leq Cd_N^2.
\end{aligned}
\]

To compare the utilities, note that the mean-value theorem gives,
for every $r,s\in\mathbb R$,
\[
|e^{-\gamma_i r}-e^{-\gamma_i s}|
\leq
\gamma_i|r-s|
\bigl(e^{-\gamma_i r}+e^{-\gamma_i s}\bigr).
\]
Conditional on the investor types, the relative wealth processes
\[
X_t^{i,N,*}
-
\frac{\theta_i}{N}\sum_{j=1}^N X_t^{j,N,*},
\qquad
X_t^{i,*}
-
\frac{\theta_i}{N}\sum_{j=1}^N X_t^{j,*},
\]
and the relative consumption processes
\[
c_t^{i,N,*}
-
\frac{\theta_i}{N}\sum_{j=1}^N c_t^{j,N,*},
\qquad
c_t^{i,*}
-
\frac{\theta_i}{N}\sum_{j=1}^N c_t^{j,*},
\]
are jointly Gaussian.
Their
means and variances are uniformly bounded because the initial
wealth, types and strategy coefficients are bounded. Their
second exponential moments are therefore uniformly bounded.

Applying the preceding inequality and the Cauchy--Schwarz
inequality to the terminal relative wealth gives a bound of
$Cd_N$ for the difference between the terminal utility terms.
For the consumption terms, Cauchy--Schwarz inequality gives
\[
\begin{aligned}
&\int_0^T
\left(
\mathbb E\Bigg|
c_t^{i,N,*}-c_t^{i,*}
-
\frac{\theta_i}{N}\sum_{j=1}^N
\bigl(c_t^{j,N,*}-c_t^{j,*}\bigr)
\Bigg|^2
\right)^{1/2}dt
\\
&\qquad\leq
\sqrt T
\left(
\mathbb E\int_0^T
\Bigg|
c_t^{i,N,*}-c_t^{i,*}
-
\frac{\theta_i}{N}\sum_{j=1}^N
\bigl(c_t^{j,N,*}-c_t^{j,*}\bigr)
\Bigg|^2dt
\right)^{1/2}
\\
&\qquad\leq Cd_N.
\end{aligned}
\]
Consequently,
\begin{equation}
\label{eq:MF-profile-payoff-stability}
\begin{aligned}
\max_{1\leq i\leq N}
\Bigg|
\mathcal J_i^N\bigl(
(\pi^{j,N,*},c^{j,N,*})_{j=1}^N;
\boldsymbol\vartheta^*
\bigr)
-
\mathcal J_i^N\bigl(
(\pi^{j,*},c^{j,*})_{j=1}^N;
\boldsymbol\vartheta^*
\bigr)
\Bigg|
\leq Cd_N.
\end{aligned}
\end{equation}

\textit{Step 3.}
Since $(\pi^{i,N,*},c^{i,N,*})_{i=1}^N$ is a Nash equilibrium
under $\boldsymbol\vartheta^*$,
\[
\mathcal V_i^N(\pi^{-i,N,*},c^{-i,N,*})
=
\mathcal J_i^N\bigl(
(\pi^{j,N,*},c^{j,N,*})_{j=1}^N;
\boldsymbol\vartheta^*
\bigr).
\]
Hence, for every $i=1,\ldots,N$ and every admissible unilateral
deviation $(\pi^i,c^i)$,
\[
\begin{aligned}
&\mathcal J_i^N\bigl(
\pi^i,c^i,\pi^{-i,*},c^{-i,*};
\boldsymbol\vartheta^*
\bigr)
\\
&\quad\leq
\mathcal V_i^N(\pi^{-i,*},c^{-i,*})
\\
&\quad\leq
\mathcal V_i^N(\pi^{-i,N,*},c^{-i,N,*})+Cd_N
\\
&\quad=
\mathcal J_i^N\bigl(
(\pi^{j,N,*},c^{j,N,*})_{j=1}^N;
\boldsymbol\vartheta^*
\bigr)+Cd_N
\\
&\quad\leq
\mathcal J_i^N\bigl(
(\pi^{j,*},c^{j,*})_{j=1}^N;
\boldsymbol\vartheta^*
\bigr)+2Cd_N,
\end{aligned}
\]
where the second and last inequalities follow from
\eqref{eq:MF-best-response-value-stability} and
\eqref{eq:MF-profile-payoff-stability}, respectively.
\end{proof}


\section{Conclusion}
\label{sec: conclusion}

We have constructed an explicit robust Nash equilibrium for a
finite-player investment--consumption game with relative performance
concerns and uncertain market coefficients. A central contribution
is the identification of explicit sufficient conditions ensuring
that the worst-case parameter selection is  deterministic and
compatible across investors trading the same asset. Our results
show that competition can change both equilibrium investment
strategies and the selected worst-case common volatility.
We have also constructed the corresponding robust mean-field
equilibrium and established convergence of the finite-player
equilibrium strategies to their mean-field counterparts. When the selected parameter profile is also
compatible with the mean-field strategies applied to finite
populations, these strategies form an asymptotic robust
$\epsilon$-Nash equilibrium.

\end{document}